\documentclass[aps,prd,eqsecnum,preprint,showpacs]{revtex4}

\usepackage{latexsym}
\usepackage{graphicx,subfigure}
\usepackage{ulem}
\usepackage{color}
\begin{document}
 
\thispagestyle{empty}
 
\title{The separate universe problem: 40 years on} 

\author{B.J.~Carr}\email[]{B.J.Carr@qmul.ac.uk}
\affiliation{ 
School of Physics and Astronomy, Queen Mary University of London, Mile End Road, London E1 4NS, UK}
\affiliation{Research Center for the Early Universe, 
University of Tokyo, Tokyo 113-0033, Japan}
\author{Tomohiro Harada}\email[]{harada-at-rikkyo.ac.jp}
\affiliation{
Department of Physics, Rikkyo University, Toshima, Tokyo 171-8501, Japan
}
\date{\today}
\begin{abstract}                
 The claim that an overdense (positive curvature) region in the early universe cannot extend beyond some maximum scale and remain part of our universe,
first made 40 years ago, 
 has recently been questioned by Kopp {\it et al.} Their analysis is elucidating and demonstrates that one cannot constrain the form of primordial density perturbations using this argument. 
However, the
 notion of a separate-universe scale still applies and it places an important upper limit on the mass of primordial black holes forming at any epoch. We calculate this scale for equations of state of the form $p = k \rho c^2$ with $-1 <k < \infty$, refining earlier calculations on account of the Kopp {\it et al.} criticisms. For $-1/3 < k < \infty$, the scale is always of order the cosmological particle horizon size, with a numerical factor depending on $k$.  
This confirms the earlier claim that a primordial black hole cannot be much larger than the particle horizon at formation. 
For $-1 < k< -1/3$, as expected for some periods in the history of the universe, 
the situation changes radically, in that
a sufficiently large positive-curvature region 
produces a baby universe rather than a black hole. There is still a separate-universe scale but the interpretation of these solutions requires care.
 \end{abstract}

\maketitle


\section{Introduction}

It is often claimed that an overdense region in an expanding
Friedmann-Robertson-Walker (FRW) background cannot extend too far, else
it would be a separate closed universe rather than a region in our universe. This was first argued by Carr and Hawking~\cite{ch1974} 
forty years ago and 
is particularly relevant to the formation of a primordial black hole (PBH) from an initial inhomogeneity because the region has to be quite close to the separate-universe limit in order to collapse against the pressure (at least in a radiation-dominated era). 

A simple way to understand this is to regard the overdense region as part of a closed ($K=+1$) FRW model in a flat ($K=0$) FRW background~\cite{carr1975}. If the size of the region is big enough to include the entire closed FRW model, then it clearly cannot be part of {\it our} universe. If one considers the region at maximum expansion, when it has some density $\rho_{m}$, then the curvature of the spacelike hypersurface must be of order $G \rho_{m}/c^2$ and so the limiting scale is of order  $c (G \rho_{m})^{-1/2}$. The density $\rho_{m}$ exceeds the density of the flat FRW background at that time by a factor $\Delta_{m}$ which depends on the equation of state. If this has the form  $p =k \rho c^2$ for some constant $k$, then $\Delta_{m} $ is a simple function of $k$ and the separate-universe scale is just the particle horizon size times a numerical factor which depends on this. The precise expression for $k>0$ was first calculated iby Harada and Carr~\cite{hc2005c}. The minimum size for a PBH is the Jeans length, which is a factor $\sqrt{k}$ smaller than the particle horizon size, so PBHs can only form over a narrow range of scales unless $k$ is very small (as in a dust era).

The above argument
is simplistic because a real overdense region would not have a constant density profile (as required for a closed FRW model) and would need to be surrounded by an underdense region (as required by the flat FRW background assumption). It is also unrealistic to model a region which evolves into a PBH as a uniform overdensity. However, at least in this context more realistic calculations -- which assume some form of density profile in the overdense region -- have been performed~\cite{nnp1978,mp2009} and these roughly support the simplistic analysis. Nobody has yet done the equivalent calculation for the separate universe problem,  even though the two problems are closely related.

Another point is that, although one can calculate the separate-universe condition by considering an overdense region at maximum expansion, this is a rather strange way of doing the calculation.
If the overdense region is a separate universe at maximum expansion, it always was and always will be a separate universe. It does not {\it evolve} into a separate universe because the topology cannot change in classical relativity. So there should be a more direct  way of doing the calculation which is time-independent and just depends upon the total energy perturbation. However, since one is dealing with fluctuations on scales initially much larger than the particle horizon, one has to be careful which measure of the perturbation is used since this is  gauge-dependent~\cite{bardeen}. The analysis in Ref.~\cite{hc2005c} uses the {\it synchronous} gauge, which is 
convenient if one considers overdense regions at maximum expansion. 

Recently, the separate-universe problem has been revisited and elucidated by Kopp {\it et al.} \cite{Kopp:2010sh} (KHW). In particular, they describe the situation in terms of the {\it curvature} perturbation 
and are not restricted to analysing the problem at the time of the overdense region's maximum expansion.  
They provide some important new conceptual insights into the problem. In particular, if one considers overdense regions of increasing size at a given epoch, they emphasize that the Misner-Sharp mass reaches a maximum and then decreases to zero at the separate-universe scale. The density fluctuation also goes to zero there, although the curvature perturbation diverges. In this sense, they claim that the separate-universe condition is naturally avoided, with the strongest constraint on the density fluctuations coming from a scale which is smaller by a factor of two. Fluctuations are described as `type I' and `type II' according to whether they are smaller or larger than this scale. 

KHW criticize the Carr-Hawking (CH) analysis, and also implicitly the
Harada-Carr (HC) analysis, on this basis.
The original CH analysis was certainly
simplistic, so it is surprising that it has gone unchallenged for 40 years. 
In particular, neither CH nor HC differentiated between type I and type II fluctuations and 
they did not appreciate that the mass of the overdense region tends to zero and the curvature perturbation diverges as one approaches the separate-universe scale. This means that one cannot constrain primordial density perturbations, as originally claimed by CH.  This is an important insight which changes one's interpretation of the problem.
The use of embedding and conformal diagrams by KHW  is also illuminating.  

Nevertheless, we have a somewhat different perspective of the problem. We agree that there are some inadequacies in the CH and HC analyses and part of the purpose of this paper is to  remedy these. However, we argue that the concept of separate-universe scale still applies and that the HC analysis provides the correct expression for this. Indeed, this expression has an important physical implication since it relates to the maximum mass of a PBH forming at any epoch in the history of the universe.  

It must be stressed that our previous analysis only applied for $k>0$ (although the expression we gave is valid for $k>-1/3$), while that of 
KHW is even more restricted, just focussing on the dust ($k=0$) and radiation ($k=1/3$) cases. On the other hand, the observed acceleration of the universe~\cite{supernova}  
can only be explained if the universe is dominated by some form of `dark energy' -- either  a quintessence field~\cite{ratra} or a perfect fluid with $p=k \rho c^2$ and $k<-1/3$ \cite{chm2010}. This would also have applied during any inflationary period in the early universe. The analysis in Ref.~\cite{hc2005c} does not apply in these situations. Indeed, the whole concept of a small initial overdensity growing until it reaches some maximum expansion fails because density fluctuations  {\it decay} with time for $k< -1/3$. So if overdense regions or black holes really can exist in such a universe, one essentially has to put them in at the outset (or at least the non-linear inhomogeneities which lead to them). 

This problem is topical because recently a one-parameter family of spherically symmetric self-similar solutions has been found which contains black holes in an asymptotically FRW background with $k<-1/3$~\cite{hmc2007,mhc2007}. This means that the black hole grows at the same rate as the cosmological horizon (interpreted as the Hubble horizon
 since there is no particle horizon in this case). What is surprising about these solutions is that the black hole cannot be too {\it large} compared to the cosmological horizon: when the ratio of the sizes is too big (above $0.7$ in the $k=-2/3$ case), one gets a cosmological {\it wormhole} (which connects two exact FRW models)
instead of a black hole~\cite{mhc09}. The transition occurs when the
black hole and cosmological apparent horizons merge. By contrast, in the
positive pressure case ($k>0$),  the black hole cannot be too {\it
small} because the solution must be supersonic everywhere.  The minimum size roughly corresponds to the Jeans condition. 

There are various other differences between these solutions. One is that, while the black hole solutions are only asymptotically quasi-FRW in the
$k>-1/3$ case, 
because there is a solid angle deficit at infinity~\cite{chm2010}, they are exactly asymptotically FRW in the $k < -1/3$ case.
Another is that in 
$k>-1/3$ case the self-similar black hole solutions tend to a `universal black hole' (in the sense that there is no cosmological particle horizon or black hole event horizon) as the parameter
which describes them (a measure of the overdensity at large distances) gets sufficiently large. 
However, there is no such limiting solution in the wormhole case and this raises the question of whether wormholes replace separate universes for $k < -1/3$ or merely represent an {\it intermediate} situation. We also need to know whether this feature is specific to the self-similar situation or applies more generally. In any case, the nature of the separate universe condition in the $k <-1/3$ situation clearly needs to be clarified. In an accompanying paper~\cite{carrharada2}, we
discuss the link between black holes and wormholes in a cosmological background in more detail.  

The plan of this paper is as follows. Sec.~II presents some general considerations, summarizing our previous approach to the separate-universe problem and emphasizing the ways in which it needs to be improved. Sec.~III covers the $k>-1/3$ case and generalizes previous treatments of the problem. Sec.~IV covers the more challenging $k \le -1/3$ case, which has not been treated before. Sec.~VI highlights some further aspects of the problem which remain to be explored. We make a more detailed comparison with the KHW analysis in Appendix A and include some mathematical details in Appendix B. 

\section{Reassessing the argument for a separate-universe scale}

In this section, we first recall the usual argument for a separate-universe scale. We then discuss various inadequacies in this argument, many of them pointed out by KHW. Finally, we present a more rigorous analysis, in which the concept of a separate-universe scale is still shown to apply. We have  no technical disagreement with the KHW analysis but we 
reformulate the problem in a way which relates more closely to the analysis of CH and HC.  

\subsection{General considerations}

We take the cosmological background to be a flat ($K=0$) FRW model with equation of state $p = k \rho c^2$,
in which the line element is given by 
\begin{equation}
ds^{2}=-dt^{2}+a^{2}(t)[dr^{2}+r^{2}(d\theta^{2}+\sin^{2}\theta)] \, .
\label{eq:metric_flat_Friedmann}
\end{equation}
Here $r$ is a dimensionless comoving radial coordinate and the scale length $a(t)$ obeys the Friedmann equation
\begin{equation}
\frac{\dot{a}^{2}}{a^2}=\frac{8\pi G\rho}{3} \, ,
\label{eq:flat_Friedmann}
\end{equation}
with a dot denoting $d/dt$. This implies the background scale length and density evolve as
 \begin{equation}
a_b \propto t^{\frac {2}{3(1+k)}} \, , \quad  \rho_b = \frac {1}{6 \pi G (1+k)^2t^2} \, ,
\label{eq:background_density}
\end{equation}
where $t$ is the time since the big bang. The size of the background particle horizon (i.e. the proper radial distance travelled by light since the big bang) is 
\begin{equation}
 R_{PH} (t) 
 = a(t)  \int^{t}_{0}  \frac {cdt}{a(t)} = \frac {3(1+k)}{1+3k}\, c t \quad (k > -1/3) \, .
\label{particle}
 \end{equation}
However, this expression diverges as $k \rightarrow -1/3$ and is inapplicable
for $k < -1/3$, so it is generally more useful to refer to the Hubble horizon:
\begin{equation}
 R_H (t)  = \frac{c}{H} = \frac{3(1+k)}{2} \, ct \quad (k > -1) 
\label{Hubble}
 \end{equation}
where $H = \dot{a}/a$. Both horizons scale as $t$ but they have a different $k$-dependence
and are equal only for $k=1/3$.
 The Hubble horizon
is smaller than the particle horizon for $k < 1/3$ and only diverges as $k \rightarrow -1$.   
Since  the mass within radius $R$ is $M= 4 \pi \rho R^3/3$, Eq.~(\ref{eq:flat_Friedmann}) implies    \begin{equation}
  \frac{ 2GM}{R c^2} = \frac {8\pi G \rho R^2}{3c^2} = \left( \frac {R}{R_H}\right)^2 \, ,
  \end{equation}
so the Hubble horizon is also equivalent to the cosmological {\it apparent} horizon.  
However, this only applies 
in a flat FRW model, when there is no ambiguity in the interpretation of the quantities $R$ and $M$. 

In order to describe the growth of an overdense region,
 one treats it as a homogeneous sphere which is part of a closed ($K=+1$) FRW model, an overdense region necessarily having positive curvature. The 
Friedmann equation then becomes
\begin{equation}
\frac{\dot{a}^{2}}{a^2}=\frac{8\pi G\rho}{3}-\frac{c^{2}}{a^{2}} \, .
\label{eq:closed_Friedmann}
\end{equation}
From Eqs.~(\ref{eq:flat_Friedmann}) and (\ref{eq:closed_Friedmann}), one
can deduce the first order equation for the evolution of a small density
perturbation
relative to the flat FRW background. This is defined as
 \begin{equation}
\delta  \equiv \frac{\rho - \rho_b}{\rho_b}
\label{defdelta}
\end{equation}
with respect to the synchronous gauge and evolves according to
 \begin{equation}
\ddot{\delta} + \frac{4 \dot{\delta}}{3(1+k)t} - \frac{2(1+3k) \delta}{3(1+k)t^2} = 0 \, .
\end{equation}
This gives a general solution
 \begin{equation}
\delta = A t^{\frac{ 2(1+3k)}{3(1+k)} } + B t^{-1} 
\label{growth}
\end{equation}
where $A$ and $B$ are constants. We took $B=0$ in our previous analysis
because only the $A$ term represents a curvature perturbation; the decaying $B$
term merely corresponds to a shift in the big bang time $t_B$, such a shift implying  
 \begin{equation}
\delta = 
\left(\frac{t}{t-t_B}\right)^2 -1 \approx \frac{2t_B}{t}  \quad (t \gg t_B) \, .
\end{equation}
However, the separate-universe condition can be modified by this effect if there is a curvature perturbation as well, so we extend the present analysis to include the $B$ term. We will want to apply Eq.~(\ref{growth}) for all values of $k$ in the range $-1$ to $+\infty$, since these are physically well motivated. Both terms decay with time for $k < -1/3$ but the first term decays more slowly for $k > -5/9$. 

\subsection{Original heuristic analysis}

We first focus on the $k > -1/3$ case and present an approximate argument of the kind used in Ref.~\cite{carr1975}. At some initial time $t_o$, when the horizon mass is $M_{Ho}$, let the density fluctuation in the positive curvature region extend to a scale $R_0$ and have amplitude $\delta_o$. If we assume that the region stops expanding when $\delta \sim 1$, then Eq.~(\ref{growth}) implies that the time  at which this happens
and its size and the 
horizon size then are given by 
\begin{equation}
t_m \sim t_o \delta _o ^{-\frac{ 3(1+k)}{2(1+3k) } } \, , \quad  R_m \sim R_o \delta _o^{-\frac{ 1}{1+3k} }  \, , \quad  R_H(t_m) \sim R_{Ho}  \delta _o^{-\frac{ 3(1+k)}{2(1+3k)} } \, .
\label{rough}
 \end{equation}
(The subscript m indicates that these conditions pertain when the radius of the region attains its `maximum'.) 
The precise type of horizon (particle or Hubble) 
is not crucial for the rough analysis presented here but it was taken to be the particle horizon in the original papers. 
Note that the terms involving $\delta_0$ in Eq.~(\ref{rough}) are large since $\delta_0$ is small and all the exponents are negative for $k > -1/3$. 

At maximum expansion, the spatial hypersurface has a radius of curvature of order $c(G \rho_m)^{-1/2}$ and the Friedmann equation implies that this is around the horizon size. Therefore the region would close upon itself and form a separate universe if it  were larger than this \cite{ch1974}. 
This situation is avoided providing   
 \begin{equation}
R_o/R_{Ho} <  \delta_o^{-1/2} \, ,
\label{roughR}
 \end{equation} 
giving a 
limit on the initial density perturbation as a function of 
scale and mass: 
 \begin{equation}
\delta_o < (R/R_{Ho})^{-2} \sim (M/M_{Ho})^{-2/3} \, .
\label{densitylimit}
 \end{equation} 
This is the limit first obtained in Ref.~\cite{carr1975}. 
On the other hand,  for collapse to a black hole,
$R_m$ must exceed the Jeans length, which is  usually taken to be $R_J \approx \sqrt{k} \, R_{H}$, so PBH formation requires
 \begin{equation}
(M/M_{Ho})^{-2/3}  > \delta_0 > k (M/M_{Ho})^{-2/3} \, 
\label{initial}
 \end{equation} 
and the density perturbation needs to be finely tuned \cite{carr1975}.

It is also useful to express this result in terms of the conditions when the positive-curvature region enters the horizon.
From Eq.~(\ref{particle}), this happens at a time and density contrast 
\begin{equation}
t_H \sim t_0 (R_0/R_{H0})^{\frac{ 3(1+k)}{1+3k} } \, , \quad  \delta_H \sim \delta_0 (R_0/R_{H0})^2 \, ,   \label{horizon}
\end{equation} 
providing $\delta_H$ is small; this will turn out to be the condition for $t_H$ to precede the maximum expansion epoch. We can now re-express Eq.~(\ref{rough}) as
\begin{equation}
t_m  \sim t_H  \delta _H ^{-\frac{ 3(1+k)}{2(1+3k) } } \, , \quad  R_m \sim R(t_H)  \delta _H^{-\frac{ 1}{1+3k} } \,  , \quad R_H(t_m) \sim R(t_H)  \delta _H ^{-\frac{ 3(1+k)}{2(1+3k) } } \, ,
\label{rough2}
 \end{equation} 
where $R(t_H)$ is the size of the overdense region when it falls within the horizon; this is not to be confused with $R_H(t_m)$, which is the horizon size at $t_m$. 
To avoid a separate universe but collapse against the pressure, one therefore requires 
 \begin{equation}
1> \delta_H > k \, ,
\label{eq:traditional}
 \end{equation} 
although this means that the density perturbation at horizon crossing is not strictly in the linear regime.
\if
$\delta_{H}$ in the above equation should merely
be regarded as an extrapolation of linear perturbation
theory to the horizon crossing time; 
Eq.~(\ref{eq:traditional}) is equivalent to Eq.~(\ref{initial})
only in so much as $\delta_{H}$ is defined in this way.
\fi
Subsequent numerical work in the radiation case ($k=1/3$) suggests that this estimate for the lower limit is quite accurate, despite all the approximations involved. While earlier work gave a value of order $0.7$ \cite{jedamsik}, more recent studies have reduced this to the range $0.3 - 0.5$ \cite{green} and calculations of PBH formation in the context of critical phenomena give the range $0.43-0.47$ \cite{musco}.  
A recent refinement of the analytical constraint (\ref{eq:traditional}), 
which is in a better agreement with the numerical 
results, gives $0.41$ in the radiation case~\cite{Harada:2013epa}.

\subsection{Problems with heuristic analysis}

In a more rigorous analysis, the significance of almost every quantity  in the above analysis ($R$, $\delta_o$, $\delta_m$, $M$, $R_H$) needs to be considered more carefully. Besides affecting the final expression for the separate-universe scale, this has important conceptual implications. Some of the points discussed below were also raised by KHW.

\begin{figure}[htbp]
\begin{center}
\includegraphics[width=0.25\textwidth]{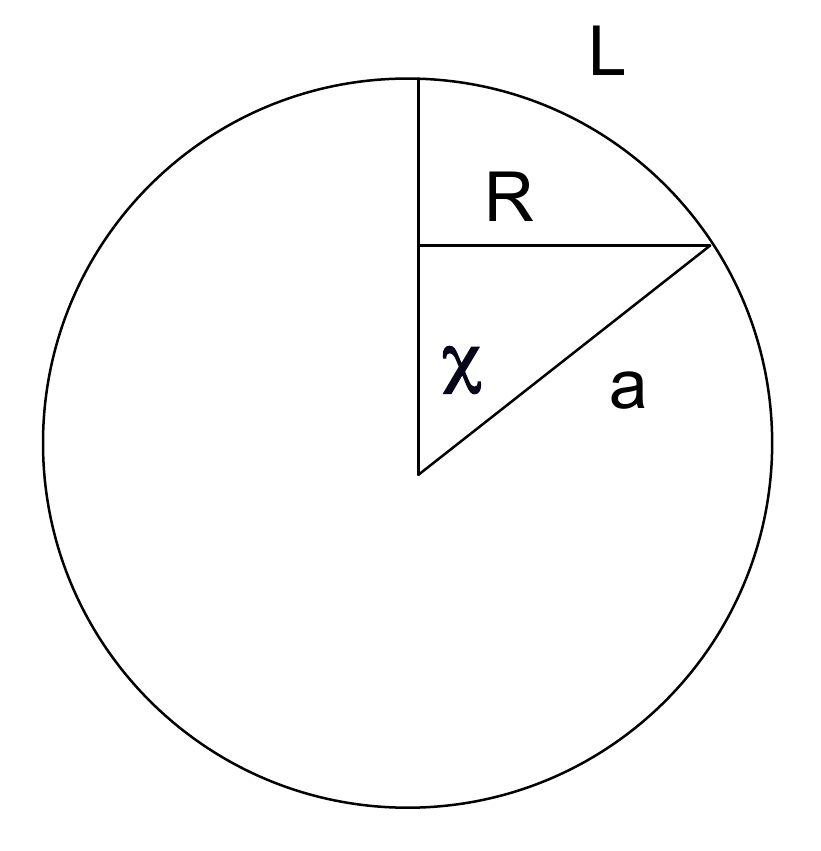}
\caption{\label{closed}
This shows the relationship between the areal radius $R$, proper radius $L$ and comoving coordinate $\chi$ in a closed FRW model at some time $t$.}
\end{center}
\end{figure}

{\it Meaning of $R$}. 
One needs to be mathematically precise in describing the geometry of an overdense region, especially when its size becomes comparable to 
the particle horizon. 
If we model it as part of a homogeneous closed FRW universe, then the line element (\ref{eq:metric_flat_Friedmann}) is replaced by
\begin{equation}
ds^{2}=-c^{2}dt^{2}+a^{2}(t) \left[\frac{dr^2}{1-K r^2} + r^2 (d\theta^{2}+\sin^{2}\theta d\phi^{2})\right]  \, ,
\label{closedmetric}
\end{equation}
where the scale length $a$ now 
gives the curvature radius of the universe and we can choose $K=1$. 
With this form of the metric, the area of the sphere of 
constant time is just $A=4 \pi a^2r^2$, so we can define an `areal radius'~\cite{hayward1996} by
\begin{equation}
R \equiv ar= (A/4\pi)^{1/2} \, .
\end{equation}
The metric can also be expressed in the form
\begin{equation}
ds^{2}=-c^{2}dt^{2}+a^{2}(t)[d\chi^{2}+\sin^{2}\chi (d\theta^{2}+\sin^{2}\theta d\phi^{2})] \, ,
\label{metric}
\end{equation}
where the dimensionless comoving radial coordinate $\chi$ is related to $r$ by $r = \sin\chi$. For fixed $t$, this gives the ``proper radius'' (i.e. the proper length from the pole 
along a great circle of the 3-sphere), which we denote by $L$. 
Thus we have
\begin{equation}
R = a \sin \chi , \quad  L = a \chi \, .
\end{equation}
While $R$ has a maximum value of $a$ at $\chi=\pi/2$ and goes to $0$ at 
$\chi=\pi$, $L$ increases monotonically until it reaches the separate-universe scale $L_{max}= \pi a$. The relationship between these quantities is illustrated in Fig.~\ref{closed}. The apparent singularity in metric (\ref{closedmetric}) at $r=1/\sqrt{K}$, corresponding to $\chi = \pi/2$, is just a coordinate effect.
The areal and proper radii are the same in a flat geometry but 
different in a closed geometry. 

{\it Meaning of M}. The expression for the mass is ambiguous in the cosmological context, especially in the positive curvature case. There are various possible definitions even in the spherically symmetric situation.
It is most convenient to use the Misner-Sharp mass, which  can be expressed in terms of the areal distance $R$ as 
\begin{equation}
M =  \frac{4 \pi \rho R^3}{3} \, .
\label{msmass}
\end{equation}
This goes to zero in the limit $\chi_a \rightarrow \pi$, essentially because the (negative) gravitational binding energy cancels the rest mass. 
As pointed out by KHW,
it is meaningless to apply Eq.~(\ref{densitylimit}) in this limit, since the right-hand side diverges. One should therefore regard the separate-universe condition as a restriction on the size rather than the mass of the overdense region. 
Although the density associated with a black hole in asymptotically flat space scales as $M^{-2}$,
which diverges as $M$ approaches zero, 
$M/R^3$ is finite in the separate-universe limit, with a value corresponding to the curvature of the background Friedmann model.

 {\it Meaning of $\delta_o$}. There are different possible measures of the initial density perturbation on scales larger than the particle horizon. The quantity $\delta_o$ is defined by Eq.~(\ref{defdelta}) as the fractional density excess relative to the flat FRW background, on the assumption that the density is homogeneous in the overdense region. However, the situation is more complicated than this, because the overdense region must be surrounded by a compensating void in order to ensure that the background is still flat. KHW therefore treat the overdense region as part of a positive-curvature FRW model with $0\le \chi\le \chi_{a}$ and let this replace the part of the 
flat FRW background with $r\le r_{b}$.
There must be a vacuum region connecting these two parts but the pressure gradient caused by the matching is assumed to be small for 
super-horizon-scale perturbations. Outside the void there is no possibility of having a separate universe, so the issue only arises on scales within the void. 
In the separate-universe limit, the mass goes to zero anyway.
KHW call perturbations `type I' for 
$0<\chi_{a}<\pi/2$ and `type II' for $\pi/2<\chi_{a}<\pi $.
Any measure of the overall density perturbation (including $\delta_o$) must depend upon 
$\chi_a$ in this simple model. 
KHW
describe the
overdensity in terms of a curvature perturbation and introduce two
measures of this -- a {\it central} volume fluctuation $\zeta (\chi_a)$
and an {\it averaged} volume fluctuation $\overline{\zeta}(\chi_a)$ --
both being nearly time-independent on scales larger than the
horizon. They also refer to a quantity $\delta(\chi_a)$, which represents the density perturbation when the overdense region enters the
horizon.  However, this is different from the (evolving) quantity $\delta$ given by Eq.~(\ref{growth}), so we write it as $\delta_{KHW}$ to avoid confusion.
As the size of the region tends to the separate-universe scale,
 $\zeta$ and $\overline{\zeta}$ 
diverge but $\delta_o$ and $\delta_{KHW}$ go to zero. 
Although the two approaches should be equivalent, 
the KHW
analysis has the advantage that it does not restrict attention to the time of maximum expansion. 

{\it Meaning of $\delta_m$}. Even if one analyses the separate-universe problem in terms of the size of an overdense region at maximum expansion, the previous heuristic approach is very imprecise. This is because the quantity $ \delta$ is not exactly $1$ at maximum expansion and the linear growth rate (\ref{growth}) only applies when $\delta$ is small. Both
these inaccuracies can be accounted for by introducing the parameter 
\begin{equation}
\Delta_m \equiv 1 + \delta_m  \equiv  \frac{\rho_m}{\rho_b(t_m)} \, ,
\label{Delta}
\end{equation}
which represents the ratio of the density in the overdense region at maximum expansion, $\rho_{m}$, to the density in the flat Friedmann background at the same time. $\Delta_m$ depends on the equation of state parameter $k$ and the problem of how to calculate this dependence is treated later. This parameter enters the calculation because
Eq.~(\ref{eq:closed_Friedmann}) implies that
the maximum scale factor is
\begin{equation}
a_{m}=\sqrt{\frac{3c^2 }{8\pi  G\rho_{m}}} 
= \Delta_m^{-1/2} R_{Hb}(t_m) \, , 
\label{eq:a_max}
\end{equation}
where $R_{Hb}(t_m)$ is the background Hubble scale at $t_m$. 
From Eq.~(\ref{Hubble}), 
the separate-universe scale at maximum expansion is
\begin{equation}
L_{max} = \pi a_m = \pi
  \Delta_m^{-1/2} R_{Hb}(t_m)\, .
\label{eq:L_max}
\end{equation}
The proper radius associated with the separate-universe
scale is always $ \pi a$, but this is time-dependent, so it is useful to specify its maximum value.

{\it Meaning of $R_H$}. The expression (\ref{eq:L_max}) was obtained in Ref.~\cite{hc2005c} and also agrees with Eq.~(16)  of KHW. However, 
there is some arbitrariness about which horizon scale should be compared to $L_{max}$.  Even if one uses the Hubble horizon rather than the particle horizon, there is still a difference between the {\it background} Hubble horizon used in Eq.~(\ref{eq:L_max}) (equivalent to the background cosmological apparent horizon) and the {\it local} Hubble horizon
(which is infinite at maximum expansion).
The ratio of the separate-universe
scale to the background Hubble scale always decreases for $k > -1/3$: it starts off large, goes as $a/t \propto
t^{-(1+3k)/3(1+k)}$ at early times and falls to a value  
less than $2$ at maximum expansion (see later).
The ratio of the separate-universe scale to the local Hubble scale
also decreases
with time but goes to zero at maximum expansion, so it is more useful to compare it to the local cosmological  apparent horizon.
 The latter is always smaller than the background Hubble horizon \cite{faraoni} and is the relevant expression so long as this lies within the positive-curvature region.
 The ratio the separate-universe scale to this horizon is exactly $\pi$ at maximum expansion \cite{carrharada2}. 

\subsection{More precise analysis}

Let us now examine how the original heuristic expression for the separate-universe scale is modified if one includes all the corrections discussed above and assumes $k>-1/3$. 
To make contact with our earlier approach, we again assume that the positive-curvature region can be described by an initial overdensity $\delta_o$, even though this is not the description used by KHW. Since $\rho \propto a^{-3(1+k)}$,
Eq.~(\ref{eq:closed_Friedmann}) can be written as
\begin{equation}
\dot{a}^2 = A a^{-(1+3k)} - B
\label{eq:a_dot}
\end{equation}
where
\begin{equation}
A \equiv \frac{8\pi}{3} G\rho_{0} a_{o}^{3(1+k)} =[c^{2}+\dot{a}_{o}^2]a_{o}^{1+3k} = \dot{a_{\rm o}}^2\Omega_{\rm o} a_{\rm o}^{1+3k}, 
 \;\;\;  B \equiv
 \dot{a}_{\rm o}^2(\Omega_{\rm o}-1) = c^2 \, .
\label{eq:A_B}
\end{equation}
Here the subscript o
 indicates some initial epoch at which the overdensity $\delta_0$ is small and 
\begin{equation}
\Omega_{\rm o} \equiv \frac{8 \pi G \rho_o}{3 H_o^2} = (1+ \delta_o) \left( \frac{H_{ob}}{H_o} \right)^2 \, .
\label{Hchange}
\end{equation}
We distinguish between the initial values of the Hubble constant in the overdense region ($H_o$) and the background  ($H_{ob}$). At a general epoch, we can express Eq.~(\ref{eq:a_dot}) in the form 
\begin{equation}
\left( \frac{H}{H_b}\right)^2 =\Omega_{M}-\Omega_{K} \, ,
\label{hubblepert}
\end{equation}
where we have introduced the epoch-dependent dimensionless parameters
\begin{eqnarray}
\Omega_{M}=\frac{8\pi G\rho}{3H_b^{2}}   ,
\quad
\Omega_{K}=\frac{c^{2}}{a^{2}H_b^{2}}  \, ,
\label{omegas}
\end{eqnarray}
From Eq.~(\ref{Hchange}), $\Omega_M(t_o) $ differs from $\Omega_o$ only by the factor $(H_{ob}/H_o)^2$.

The decomposition on the right-hand-side of Eq.~(\ref{Hchange}) is gauge-dependent. For $k>-1/3$, we can always choose a (uniform Hubble) gauge in which $H_{ob} = H_o$ initially. In this case, the ratio of the local density to the background density, $\Delta(t)$, always exceeds unity
and initially 
\begin{equation}
\Omega _o =  \Delta(t_o) = 1 +  \delta _o  \, .
\label{omegadelta}
\end{equation}
However, another choice of 
gauge  would be possible
where this need not be the case. Indeed, for $k<-1/3$, we will see later that a positive-curvature region collapses from a dispersed state and then bounces into expansion, so it is necessarily {\it underdense} relative to the flat FRW background close to the big bang.

Since the expression for $B$ in Eq.~(\ref{eq:A_B}) implies
\begin{equation}
H_o = \frac{c}{a_o \sqrt{\Omega_o-1}} \, ,
\label{curenthubble} 
\end{equation}
the initial size of the region is given by
\begin{equation}
\frac{L_o}{R_{Ho}} = \frac{\chi_a}{\sqrt{\Omega _o-1}},  \quad 
\frac{R_o}{R_{Ho}} = \frac{\sin \chi_a}{\sqrt{\Omega _o-1}}   \, .
\label{initialL}
 \end{equation}
Here $R_{Ho}$ is the initial {\it local} Hubble scale, which differs from the initial {\it background} Hubble scale $R_{Hob}$ by the factor $H_{ob}/H_o$. 
Putting $\dot{a}=0$ in Eq.~(\ref{eq:a_dot}) gives the size of the region at maximum expansion: \begin{equation}
L_m = a_{m} \chi_a =  a_{o} \chi_a \left( \frac{\Omega_o}{\Omega_o -1}
			       \right)^{\frac{1}{1+3k}} = L_o\left(
			       \frac{\Omega_o}{\Omega_o -1}
						       \right)^{\frac{1}{1+3k}} \, ,
\end{equation}
\begin{equation}
R_m = a_{m} \sin\chi_a =  a_{o} \sin\chi_a \left( \frac{\Omega_o}{\Omega_o -1}
			       \right)^{\frac{1}{1+3k}} = R_o\left(
			       \frac{\Omega_o}{\Omega_o -1}
						       \right)^{\frac{1}{1+3k}} \, .
\label{Rmax}
\end{equation}
Since the overdensity at maximum expansion can be expressed as 
\begin{equation}
\Delta_m = \frac{\rho_m}{\rho_o}\frac{\rho_{bo}}{\rho_{bm}}\frac{\rho_o}{\rho_{bo}} =  \frac{\rho_m}{\rho_o} \left( \frac{t_m}{t_o}\right)^2  \Omega_o  \left( \frac{H_o}{H_{ob}} \right)^2 \, ,
\label{deltamax}
\end{equation}
we can write the time of maximum expansion as 
\begin{equation}
t_m =
t_o \Delta_m^{1/2} \Omega _o^{\frac{1}{1+3k}} (\Omega _o-1)^{-\frac{ 3(1+k)}{2(1+3k) }} 
 \left( \frac{H_{ob}}{H_o} \right) \, .
\end{equation}
Condition (\ref{roughR}) 
is then replaced by
 \begin{equation}
\frac{L_o}{R_{Ho}} < \frac{\pi}{\sqrt{\Omega _o-1}} \,  ,
\label{maxscale}
\end{equation}
which corresponds to putting $\chi_a = \pi$ in Eq.~(\ref{initialL}). From Eq.~(\ref{curenthubble}) this is just equivalent to the condition $L_o < \pi a_o$, as expected. In the uniform-Hubble gauge, the right-hand-side of Eq.~(\ref{maxscale}) can be written as $\pi \delta_o^{-1/2}$, so this might be regarded as a more precise version of the heuristic limit (\ref{roughR}). From Eq.~(\ref{initialL}) there is also a purely geometrical upper limit,
\begin{equation}
\frac{R_o}{R_{Ho}} \le \frac{1}{\sqrt{\Omega _o-1}} \,  ,
\end{equation}
with equality for $\chi_a = \pi/2$, corresponding to the condition $R_o < a_o$, but this is distinct from the separate-universe constraint since $R_o$ goes to zero at $\chi_a = \pi$. 

\begin{figure}[htbp]
\begin{center}
\includegraphics[width=0.5\textwidth]{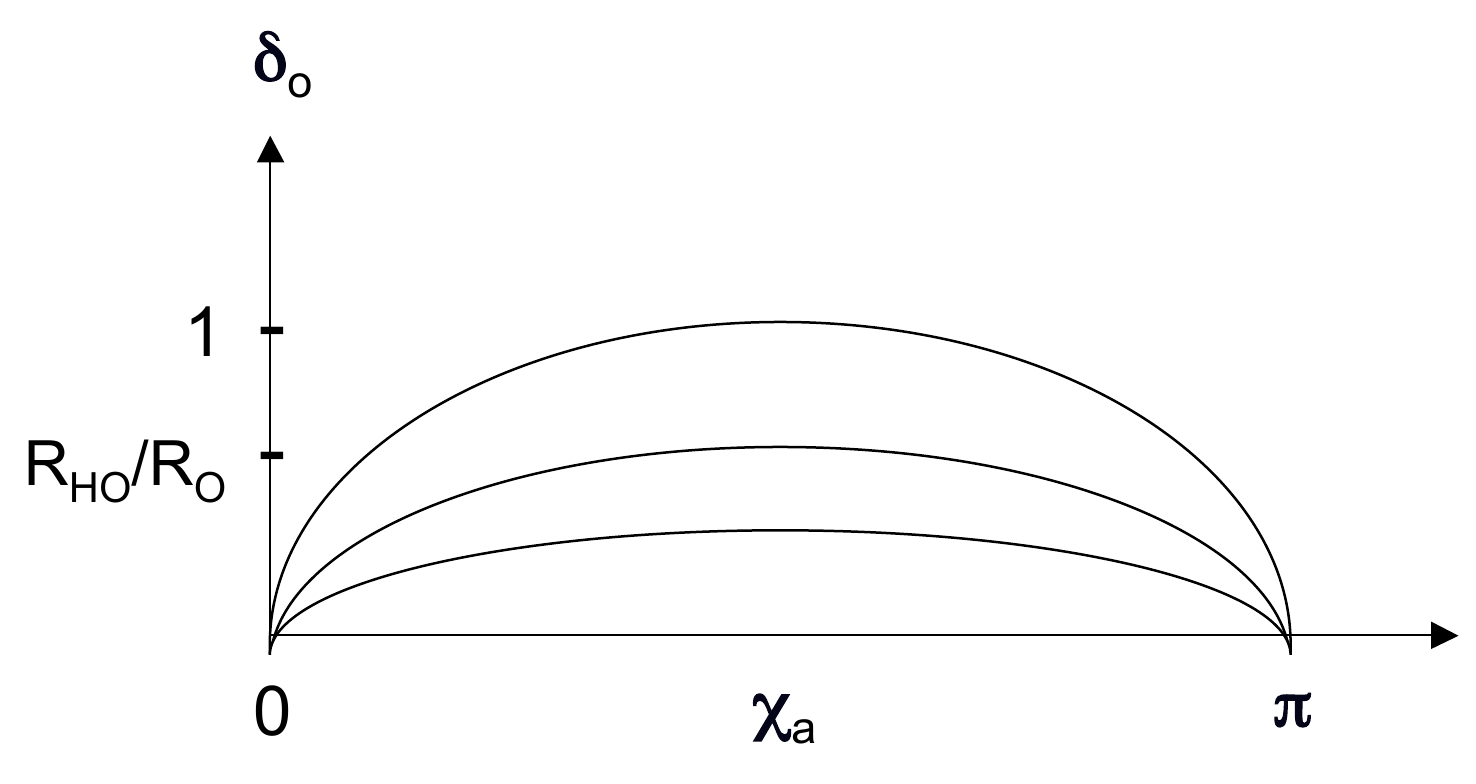}
\caption{\label{delta}
This shows the dependence of the overdensity  $\delta_0$ on $\chi_a$. It is zero at $\chi =0$ and $\chi = \pi$ and has a maximum value of $R_{Ho}/R_o$, which is $1$ if the horizon epoch is taken as the initial time.}
\end{center}
\end{figure}

If we adopt Eq.~(\ref{omegadelta}), corresponding to $H_{ob} = H_o$, and use the relation $R_o = a_o \sin \chi_a$, 
we can interpret Eq.~(\ref{curenthubble}) as specifying $\delta_o$ as a function of $\chi_a$:
\begin{equation}
\delta_o = (R_{Ho}/R_o)^2 \sin^2 \chi_a,
\label{deltao}
\end{equation}
this being zero  at $\chi = 0$ and $\chi = \pi.$ In this case, we can regard the factor $R_{Ho}/R_o$ as representing the dependence of $\delta_o$ on $t_o$ for given $\chi_a$. As illustrated in Fig. \ref{delta}, this factor increases with $t_o$, starting small for $t_0 \ll t_H$ and becoming $1$ at horizon-crossing ($t_o = t_H$). This contrasts with the original CH analysis, where the parameter $\delta_o$ is assumed independent of $\chi$, and with the KHW analysis, where  the parameter $\zeta$ is constant in time until gradients build up after horizon-crossing.

We can use Eq.~(\ref{omegadelta}) to express all our conditions in terms of  $\delta_o$.
In particular, Eq.~(\ref{rough}) is replaced by the {\it exact} relations
\begin{equation}
t_m = t_o \Delta_{m}^{1/2}  
\delta_o^{-\frac{ 3(1+k)}{2(1+3k) }} (1+ \delta_o)^{\frac{ 1}{1+3k }} ,  \quad  
L_m = L_0 \left( \frac{\delta _o}{1+ \delta
		     _o}\right)^{-\frac{ 1}{1+3k} }  , 
\quad   R_{Hb}(t_m) = R_{Ho}  \Delta_{m}^{1/2} 
\delta_o^{-\frac{ 3(1+k)}{2(1+3k) }} (1+ \delta_o)^{\frac{ 1}{1+3k }}   \, ,
\label{exact1}
\end{equation}
and condition (\ref{densitylimit}) becomes
 \begin{equation}
\delta_o
< \left( \frac{ \pi R_{Ho}}{L_o} \right)^2 \, .
\label{eq:separate_universe_exact}
 \end{equation}
This differs from the original condition because it contains $L_0/\pi$ rather than $R_0$ and because it is not expressed in terms of $M$.
More importantly, Eq.~(\ref{deltao}) shows that this condition reduces to
 \begin{equation}
 L_o/R_o < \pi/ \sin \chi_a \, .
 \end{equation}
Since $R_o = a_o \sin \chi_a$   and $L_o = a_o \chi_a$ with $\chi_a < \pi$, this constraint is  {\it necessarily} satisfied. Thus one still has a maximal proper length-scale for a fluctuation in our universe but there is no constraint on the fluctuation amplitude itself other than the purely geometrical condition $R_o < a_o$. In this sense, we agree with the conclusion of KHW that condition (\ref{densitylimit}), even when expressed correctly, is not important.

Expressing the separate-universe condition in terms of the overdensity when the region falls within the horizon (i.e. the analogue of  Eq.~(\ref{rough2})) is problematic because we have seen that there is some ambiguity in what is meant by horizon-crossing. Since the areal radius rather than the proper radius is relevant for embedding a positive-curvature FRW model 
in a flat FRW background,
we will regard horizon-crossing as occurring when the 
areal radius of the overdense region matches the background Hubble horizon. 
This has the important implication that a type II fluctuation enters the horizon very early, which means that a type II PBH still has a size comparable to the horizon at formation. 

Applying Eq.~(\ref{deltao}) 
when $R_o = R_{Ho}$ implies 
\begin{equation}
 \delta _H=\sin^{2}\chi_{a} \, .
\label{deltah}
 \end{equation}
However, $R_{Ho}$ is the local Hubble scale and this is different from both the local cosmological apparent horizon scale $R_{CAHo}$ and the background Hubble scale $R_{Hob}$. One can show that these quantities are related by \cite{carrharada2}
\begin{equation}
R_{CAHo} = \Omega_{Mo}^{-1/2}R_{Hob} = \Omega_o^{-1/2}R_{Ho}  \, , 
\end{equation}
so Eq.~(\ref{deltao}) gives
\begin{equation}
 \delta _H= \frac{1}{2} (\sqrt{1+4\sin ^2 \chi_a} -1) \approx  \sin^{2}\chi_{a} (1 -  \sin^{2}\chi_{a}) \, .
 \end{equation}
This might be compared to the KHW expression in the radiation case,
\begin{equation}
\delta_{KHW} = \frac{1}{16} \sin^2\chi_a(8 + \sin^2 \chi_a)  \, ,
\end{equation}
 which differs by a factor $(8+ 9\sin^2 \chi_a)/16$ for $\chi_a \ll 1$.
(A simple analytic formula is only possible in this case.) The difference arises because they choose a gauge in which the time rather than the Hubble rate in the $K=0$ and $K=+1$ FRW models are identified (i.e. they use a synchronous rather than uniform Hubble gauge). 
Both expressions go to $0$ as $\chi_a \rightarrow \pi$ and
Eq.~(\ref{rough2}) is replaced with
\begin{equation}
t_m = t_H \Delta_{m}^{1/2}  \delta _{H}^{-\frac{ 3(1+k)}{2(1+3k) } } (1+ \delta_H)^{\frac{ 1}{1+3k }} \, , \quad  R_m = R_H \delta _H^{-\frac{ 1}{1+3k} } \,  , \quad R_{Hb}(t_m) = R(t_H) \Delta_{m}^{1/2} \delta_{H}^{-\frac{ 3(1+k)}{2(1+3k) } }(1+ \delta_H)^{\frac{ 1}{1+3k }}  \, .
\label{exact2}
 \end{equation}
The limit Eq.~(\ref{eq:separate_universe_exact}) 
becomes
\begin{equation}
 \delta _H
 \le 
 \pi^2 \, . 
\end{equation}
but again this condition is ensured by  Eq.~(\ref{deltah}), so nothing new is learnt.  
The only disagreement with KHW concerns the divergence of the curvature perturbation at $L_{max}$. Since $\zeta$ cannot exceed infinity, KHW interpret this to mean the separate-universe condition cannot be achieved, whereas we interpret it to mean that $\zeta$ necessarily diverges!

Note that PBH formation is only possible if $L_m$ exceeds the {\it local} Jeans length, which is a factor $\Delta_{m}^{1/2}$ smaller than the background Jeans length. However, the condition for PBH formation is still given by Eq.~(\ref{initial}) because this cancels the factor of  $\Delta_{m}^{1/2}$  for $R_{Hb}(t_m)$ in Eq.~(\ref{exact2}). It is interesting that the ratio of the Jeans scale to the separate-universe scale can exceed $1/2$ for some values of $k$, implying that a PBH derives necessarily from a type II perturbation. The expression for the Jeans length adopted in Ref.~\cite{hc2005c} is 
 \begin{equation}
R_J = \frac{4 \pi \sqrt{k}}{5+9k} c H^{-1} = 2 \pi \sqrt{k} \left( \frac{1+3k}{5+9k}\right)  R_{PH} \, ,
  \end{equation}
where the last coefficient scales as $\sqrt{k}$ for both $k \ll 1$ and  $k \gg 1$. The exact expression is intriguingly close to half the separate-universe scale for all values of  $k$ between  $0.1$ and $1$. 
 
As pointed out by KHW, these considerations  also have implications for the probability of PBH formation. One often assumes \cite{carr1975} that the horizon-epoch density fluctuations $\delta_H(M)$ have a Gaussian distribution as a function of $M$ with some root-mean-square value $\epsilon(M)$. The probability of PBH formation for a given value of $M$ then goes as
  \begin{equation}
  P(M) \sim \int_{\delta_{min}(M)}^{\delta_{max}(M)} \exp \left [-\left(\frac{\delta_H}{ \epsilon(M)}\right)^2 \right] d \delta_H \, ,
  \end{equation}
 where $\delta_{max}(M)$ is associated with the separate-universe condition. With the CH argument, the precise form of the upper limit does not matter and it might as well be taken to be infinity because of the exponential term. However, if $\delta \rightarrow 0$ at the separate-universe scale, the distribution cannot be exactly Gaussian and it is misleading to regard the upper limit as a function of $M$ at all. As stressed by KHW, only the curvature perturbation $\zeta$ can be expected to be Gaussian and in this case the upper limit becomes infinite. Perhaps the most important aspect of the separate-universe calculation is that it implicitly specifies any upper limit on the size of a PBH forming at any epoch, this being half the separate-universe scale. Note that in the inflationary scenario there may also be an intrinsic non-Gaussianity in the distribution of $\delta_H$~\cite{hidalgo} but we neglect this effect here.

\subsection{Extending the analysis to the $k < -1/3$ case}

In the $k \le -1/3$ case, discussed in detail in Sec~IV, the situation is fundamentally different in a number of respects. First, Eq.~(\ref{growth}) implies that density inhomogeneities do not grow with time, so  the concept of a slightly overdense region reaching some maximum expansion is inapplicable. 
Second, 
there is no particle horizon in this case, so one must use the  Hubble horizon
instead. 
Third, the significance of the factor $\Delta_{m}$ is different, since $R$ may be momentarily static {\it before} $t_0$ and corresponds to a minimum rather than a maximum.
The behaviour of the relativistic parameter $2GM/c^2R \propto a^{-(1+3k)}$ is similar in both the  $k > -1/3$ and  $k < -1/3$ cases: it first decreases from a large value, reaches a minimum which is necessarily less than $1$ and then increases to a large value. However,  the physical interpretation is different: 
for  $k > -1/3$, any positive-curvature region first goes outside a local cosmological apparent horizon and later falls within a black hole apparent horizon (unless the pressure stops the collapse); for  $k <  -1/3$, it first goes outside a  white hole apparent horizon and later falls within a local cosmological apparent horizon. We discuss the significance of this further in Ref.~\cite{carrharada2}.

\section{Separate universe condition for  $k>-1/3$}

If we assume the overdense region is part of a closed FRW
model with metric~(\ref{metric}), then the largest proper radius of a 3-sphere at maximum expansion is $\pi$ times the curvature radius $a_{m}$ given by Eq.~(\ref{eq:a_max}). 
Therefore it only remains to determine the  overdensity at maximum expansion $\Delta_{m}$.
In the dust case, this is well known to be $(3\pi/4)^2$ and this result was generalized to the $1 > k>0$ case in Ref.~\cite{hc2005c}. We reproduce our earlier argument here but extend it to include the $-1/3 < k < 0$ case and also allow for a variation in the big bang time.  
Some technical mathematical aspects of the calculation can be found in Appendix B.

For general $k$, the evolution of $a$ is given by Eq.~(\ref{eq:a_dot}).
By defining a new scale factor $b$ and a new time coordinate $\tau$ such that
\begin{equation}
b =  a^{1+3k}, \;\;\; d\tau = (1+3k) b^{\frac{3k}{1+3k}} dt, 
\label{eq:b_tau}
\end{equation}
we can transform Eq.~(\ref{eq:a_dot}) into the dust form:
\begin{equation}
\left(\frac{db}{d\tau}\right)^2 = \frac{A}{b} - c^2.
\label{eq:dbdtau}
\end{equation}
Note that $b$ and $\tau$ have dimensions $[L^{1+3k}]$ and $[TL^{3k}]$, respectively. This has the parametric solution
\begin{equation}
b = \frac{1}{2} b_{\rm m}(1-\cos \eta), \;\;\; \tau =\frac{1}{\pi}
 \tau_{\rm m} (\eta - \sin \eta), \;\;\;  \tau_m = \frac{b_m \pi}{2c} \, ,
\label{eq:b_eta_tau_eta}
\end{equation}
where $\eta = \pi$ corresponds to the epoch of maximum expansion  (indicated by subscript m).  
Equations~(\ref{eq:b_tau}) and (\ref{eq:dbdtau}) then imply
\begin{eqnarray}
t &=& \frac{1}{1+3k} \int^{\tau}_{0}b^{-\frac{3k}{1+3k}}d\tau \nonumber \\
&=& \frac{1}{1+3k} 
\left(\frac{b_{\rm m}}{2}\right)^{-\frac{3k}{1+3k}}
\left(\frac{\tau_{\rm m}}{\pi}\right)
\int^{\eta}_{0} (1-\cos \eta)^{\frac{1}{1+3k}} d\eta .
\label{eq:t_eta}
\end{eqnarray}
This assumes that the overdense region has the same `big bang' time as the background; we will examine the consequences of dropping this assumption later.  Equations~(\ref{eq:a_dot}), (\ref{eq:A_B}) and (\ref{eq:b_tau}) give
\begin{equation}
\frac{a_{\rm m}}{a_{\rm o}} = \left(\frac{\Omega_{\rm o}}{\Omega_{\rm o} -1}\right) ^{\frac{1}{1+3k}},\;\;\;
\frac{b_{\rm m}}{b_{\rm o}} =\left( \frac{\Omega_{\rm o}}{\Omega_{\rm o} -1}\right) ,
\label{maxsize}
\end{equation}
so we have
\begin{equation}
\frac{\rho_{\rm m}}{\rho_{\rm o}} = \left(\frac{a_{\rm m}}{a_{\rm o}}\right)^{-3(1+k)} =  \left(\frac{\Omega_{\rm o} -1}{\Omega_{\rm o}}\right) ^{\frac{3(1+k)}{1+3k}} .
\end{equation}
Since Eq.~(\ref{eq:background_density}) implies that the background density at $t_m$ is given by
\begin{equation}
\frac{\rho_{\rm bm}}{\rho_{\rm bo}} = \left(\frac{t_o}{t_{\rm m}}\right)^2,
\end{equation}
where $t_0$ and $t_{\rm m}$ are given by Eq.~(\ref{eq:t_eta}) with $\eta=\pi$ and $\eta=\eta_{0}$, respectively, Eq.~(\ref{deltamax}) can be written as
\begin{equation}
\Delta_{m} 
= \left(\frac{\Omega_{\rm o} -1}{\Omega_{\rm o}}\right)
 ^{\frac{3(1+k)}{1+3k}} 
\left[ \frac{\int^{\pi}_{0} (1-\cos \eta)^{\frac{1}{1+3k}}
 d\eta}{\int^{\eta_{\rm o}}_{0} (1-\cos  \eta)^{\frac{1}{1+3k}}
 d\eta}\right]^2  \Omega_{\rm o}  \left( \frac{H_o}{H_{ob}} \right)^2\, .
\label{eq:Delta_Omega_eta}
\end{equation}

From Eq.~(B7) of Appendix B, the top integral in Eq.~(\ref{eq:Delta_Omega_eta}) can be expressed as
\begin{equation}
\int^{\pi}_{0} (1-\cos \eta)^{\frac{1}{1+3k}} d\eta=
2^{\frac{1}{1+3k}}\sqrt{\pi}\; 
\frac{\Gamma \left( \frac{3(1+k)}{2(1+3k)}\right)}{ 
\Gamma \left( \frac{2+3k}{1+3k}\right )} \,  ,
\label{colltime}
\end{equation}
where $\Gamma$ denotes the gamma function.
From Eqs~(\ref{eq:b_eta_tau_eta}), (\ref{maxsize}) and (B8), the integral limit on the bottom is
\begin{equation}
\eta_{\rm o} = \cos^{-1}\left(\frac{2-\Omega_{\rm o}}{\Omega_{\rm o}}\right) \approx
 2\sqrt {\frac{\Omega_{\rm o} -1}{\Omega_{\rm o}}} \, \left[ 1 + \frac{1}{6} \left( \frac{\Omega_{\rm o} -1}{\Omega_{\rm o}} \right) \right] \, ,
\label{eq:eta_Omega}
\end{equation}
where we assume $\Omega_{0} -1 \ll 1$ in the last approximation, so that $\eta_{\rm o}$ is also small. Using Eq.~(B9), the lower integral 
in Eq.~(\ref{eq:Delta_Omega_eta}) can then be approximated by
\begin{equation}
\int^{\eta_{\rm o}}_{0} (1 - \cos \eta)^{\frac{1}{1+3k}} d\eta 
\approx 2^{\frac{2+3k}{1+3k}}\left[\frac{1+3k}{3(1+k)}\right] \left(\frac{\Omega_{\rm o} -1}{\Omega_{\rm o}}\right) ^{\frac{3(1+k)}{2(1+3k)}}  \left[ 1 + \frac{3(1+k)}{2(5+9k)} \left( \frac{\Omega_{\rm o} -1}{\Omega_{\rm o}} \right) \right] \, .
\label{initialtime}
\end{equation}
The first term involving
$\Omega_{\rm o}$ cancels the equivalent term in Eq.~(\ref{eq:Delta_Omega_eta}) but a dependence on $\Omega_{\rm o}$ still appears as a small correction in the second term. This gives
\begin{equation}
\Delta_{m}=\frac{\pi \alpha}{4}\left[\frac{3(1+k)}{1+3k}\right]^{2}
\frac{\Gamma\left( \frac{3(1+k)}{2(1+3k)}\right)^{2}}
{\Gamma \left( \frac{2+3k}{1+3k}\right )^{2}} = \frac{\pi \alpha \, \Gamma\left( \frac{5+9k}{2(1+3k)}\right)^{2}} 
{\Gamma \left( \frac{2+3k}{1+3k}\right )^{2}}  \, 
\end{equation}
where 
\begin{equation}
\alpha \approx \left[ 1 + \frac{2(1+3k)}{5+9k} \left( \frac{\Omega_{\rm o} -1)}{\Omega_{\rm o}} \right) \right] \Omega_{\rm o} \left( \frac{H_o}{H_{ob}} \right)^2
\end{equation}
is close to $1$ and gives only a weak dependence on $\Omega_o$. 
As $k$ decreases from $\infty$ to $-1/3$, $\Delta_m$ increases from $(\pi/2)^2 \approx 2.3$ to $\infty$, passing through $4$ at $k=1/3$ and $(3\pi/4)^2 \approx 5.3$ at $k=0$.

\begin{figure}[htbp]
\begin{center}
\includegraphics[width=0.4\textwidth]{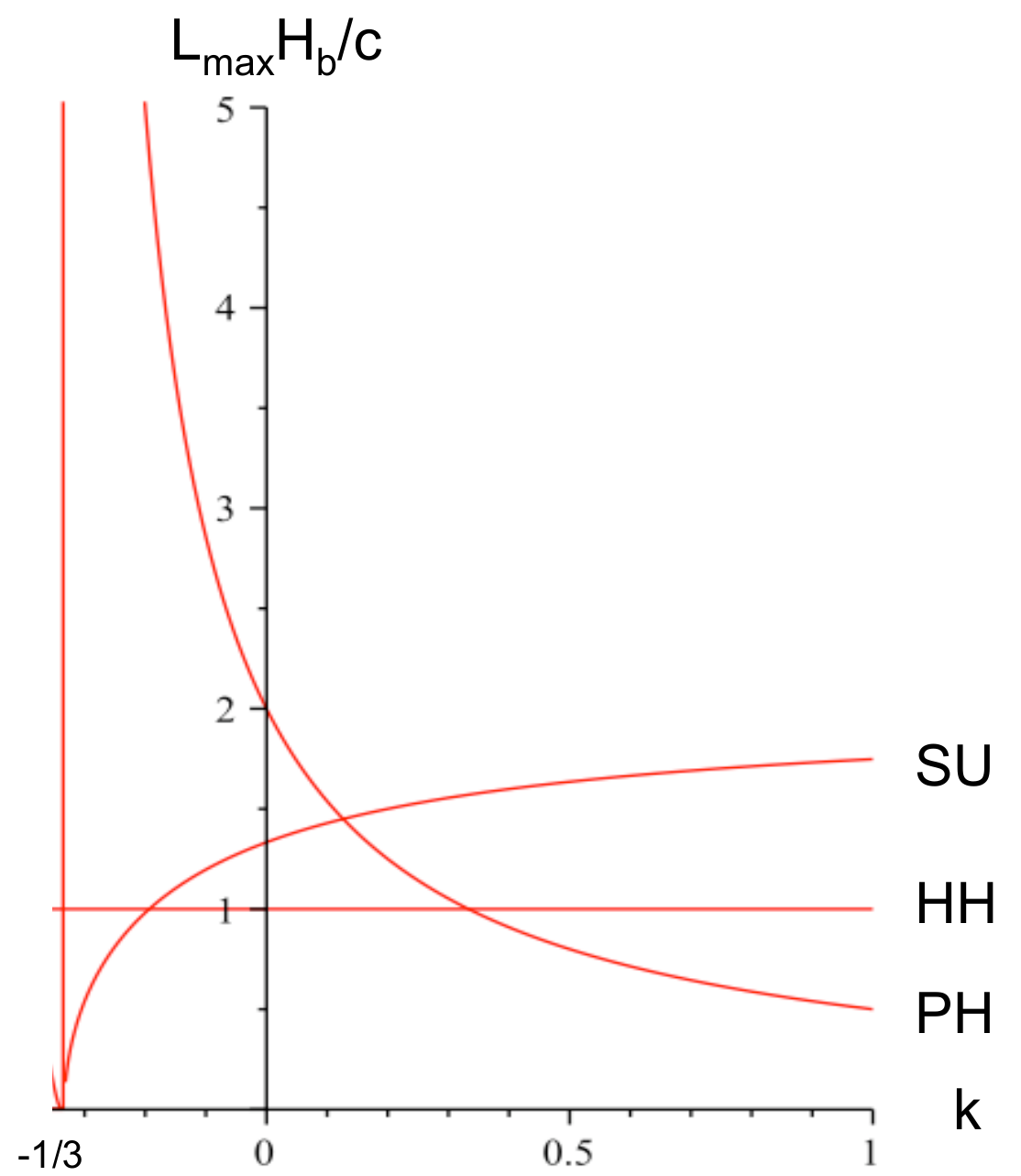}
\caption{\label{nob}
This shows the separate-universe scale (SU), Hubble horizon scale (HH)
 and particle horizon scale (PH)  in units of the background Hubble
 length as a function of the equation of state parameter $k$ when the time of the big bang is unperturbed. The maximum size of a PBH is half the SU size.}
\end{center}
\end{figure}

Equation~(\ref{eq:L_max}) now gives the separate
universe scale in units of the background Hubble scale $c/H_{bm}$ at maximum expansion: 
\begin{equation}
c^{-1}H_{bm}L_{\rm max}=\pi\Delta_{m}^{-1/2}= 2 \sqrt{\frac{\pi}{ \alpha} }\, \frac{(1+3k)}{3(1+k)} \frac{
\Gamma \left( \frac{2+3k}{1+3k}\right )}{ \Gamma \left( \frac{3(1+k)}{2(1+3k)}\right)} 
=  \sqrt{\frac{\pi}{ \alpha} }\,  \frac{ \Gamma \left( \frac{2+3k}{1+3k}\right )}{ \Gamma \left( \frac{5+9k}{2(1+3k)}\right)} \, ,
\label{maximum}
\end{equation}
the factor of $5+9k$ in the last term relating to the ratio of the exponents in Eq.~(\ref{growth}). This expression is plotted in Fig.~\ref{nob}. In a dust universe ($k=0$), this gives $4/3$. In a 
radiation universe ($k=1/3$), it gives $\pi/2$. In a (possibly unphysical) 
universe with $k=\infty$, it gives $2$.
In the limiting case $k=-1/3$, it goes to zero but we discuss this more carefully later. Note that $L_{\rm max}$ exceeds the cosmological particle horizon for 
$k\agt 0.126$ and the cosmological apparent horizon (or Hubble horizon) for 
$k\agt - 0.2$.
For comparison, the maximum areal radius of an overdense region, associated with the maximum size of a PBH,  
is smaller by a factor of $\pi$.
Although we have presented our analysis in terms of the evolution of an overdense region, it should be stressed that  a region which is
a separate universe at maximum expansion is {\it always} a separate universe. 

\begin{figure}[htbp]
\begin{center}
\includegraphics[width=0.8\textwidth]{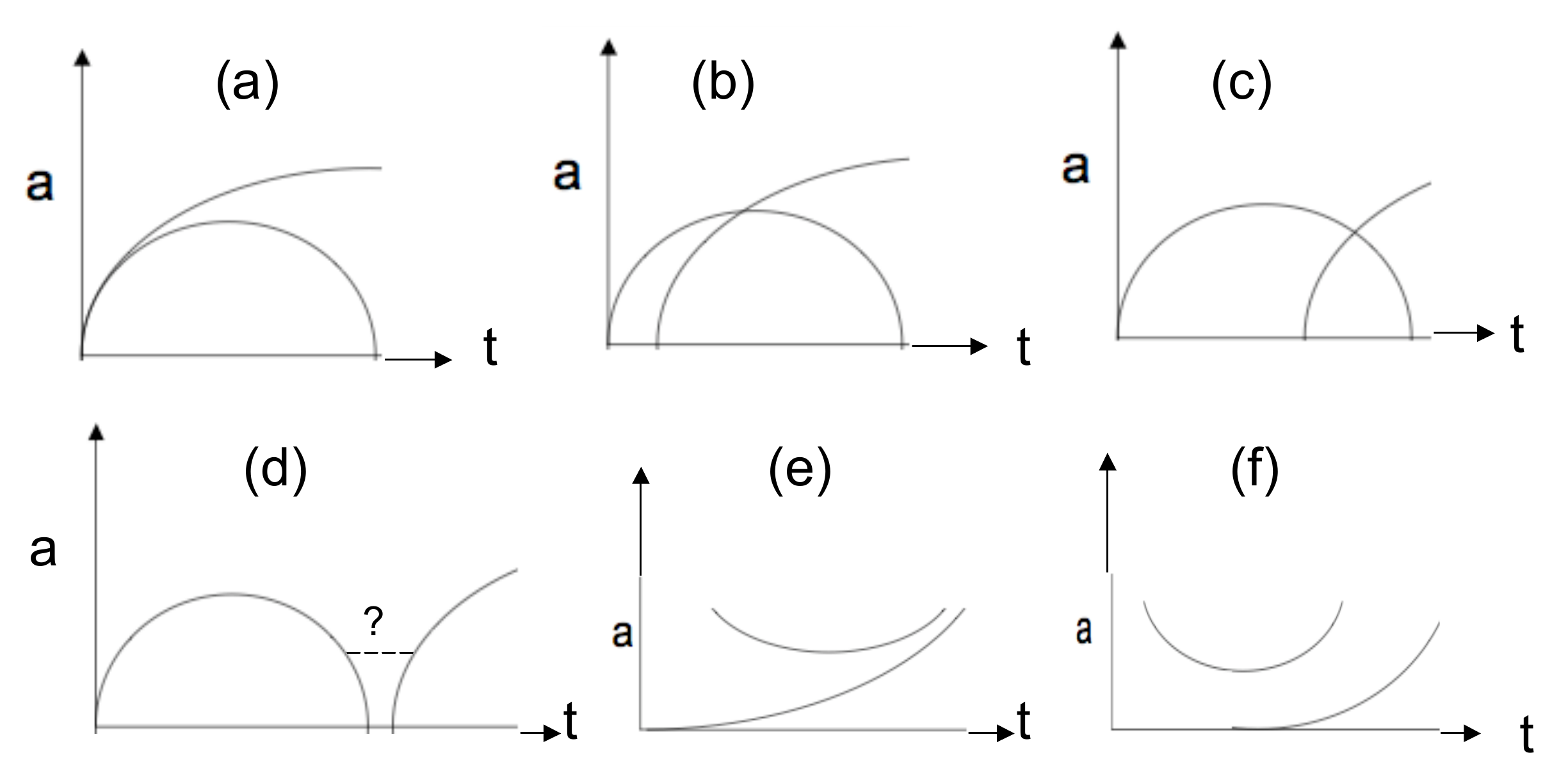}
\caption{\label{delayed}
This shows the evolution of the scale factor for the positive curvature region and background for $k>-1/3$ (cases a to d) and $k<-1/3$ (cases e and f) with various big bang times. }
\end{center}
\end{figure}

\begin{figure}[htbp]
\begin{center}
\includegraphics[width=0.5\textwidth]{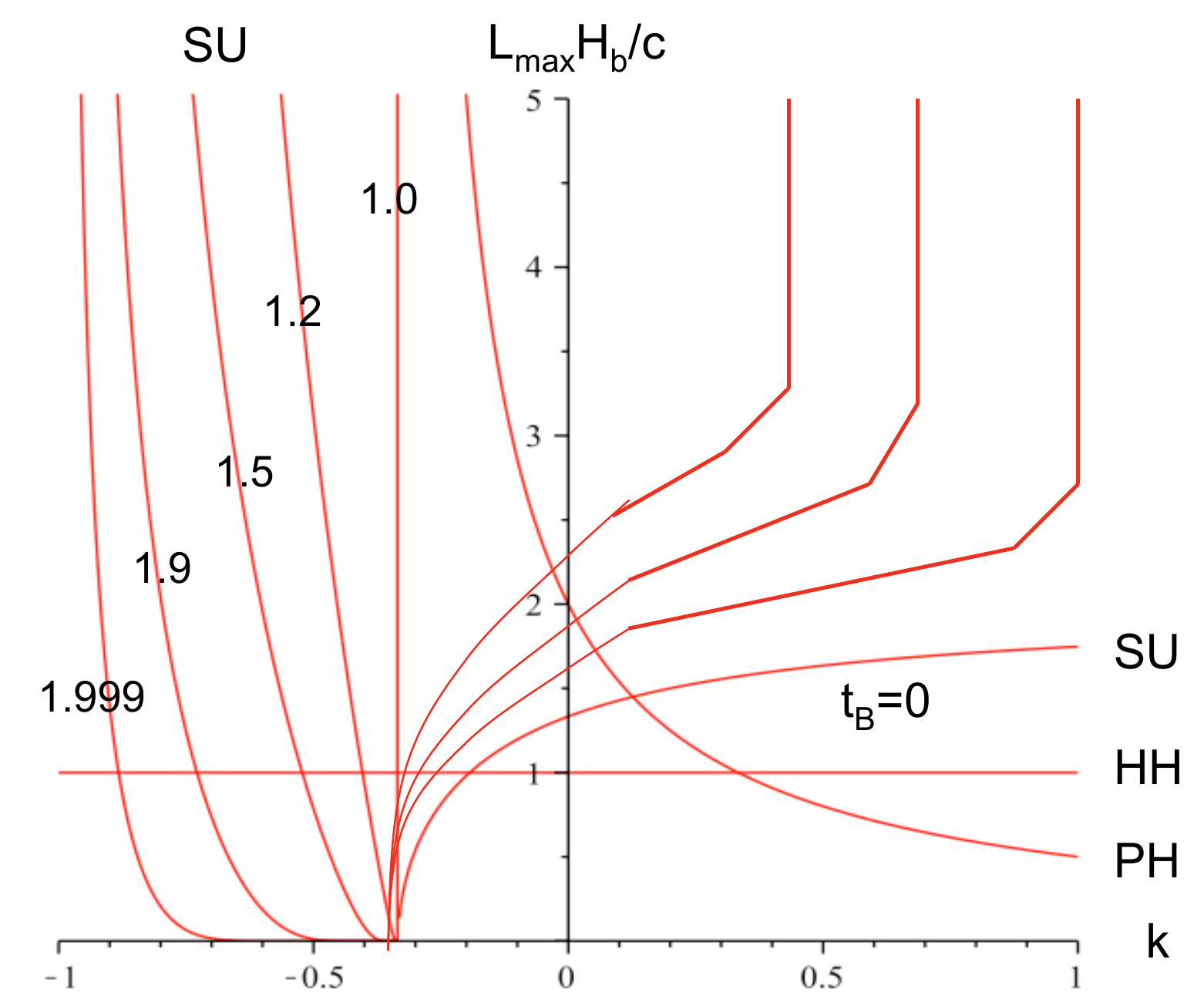}
\caption{\label{fg:horizon}
As in Fig.~\ref{nob} but allowing for a perturbed big bang time. 
 The SU
 curves depend on the big bang delay in the background universe relative
 to the positive curvature region. For $k>-1/3$, the solid line
 corresponds to no delay ($t_B=0$) and the other lines correspond to increasing
 delay. 
  For $k < -1/3$, the curves are labelled  by
 $\eta_B/\pi$ where $\eta$ is conformal time. }
\end{center}
\end{figure}

In principle, the perturbed region could have a big bang time $t_B$ which is different from 
the background universe. This corresponds to the decaying term $Bt^{-1}$ in Eq.~(\ref{growth}). One has various possible situations as summarized in Fig.~\ref{delayed}. The previous analysis applies when $t_B=0$ (case a). Providing $t_B < t_m$, so that the big bang in the background occurs before maximum expansion in the overdense region (case b), we can choose $t_0$ to be after $t_B$ and then replace Eq.~(\ref{eq:Delta_Omega_eta}) with 
\begin{equation}
\Delta_m 
= \left(\frac{\Omega_{\rm o} -1}{\Omega_{\rm o}}\right)
 ^{\frac{3(1+k)}{1+3k}} 
\left[ \frac{\int^{\pi}_{0} \xi (1-\cos \eta)^{\frac{1}{1+3k}}
 d\eta - t_B }{\int^{\eta_{\rm o}}_{0} \xi (1-\cos  \eta)^{\frac{1}{1+3k}}
 d\eta - t_B} \right]^2 \Omega_o  \left( \frac{H_o}{H_{ob}} \right)^2 \, .
\label{messy}
\end{equation}
Here the timescale $\xi$ is given by
\begin{equation}
\xi =   \frac{1}{c \, (1+3k)}  \left(\frac{b_{\rm m}}{2}\right)^{\frac{1}{1+3k}}
= \frac{2^{-\frac{1}{1+3k}}}{(1+3k)H_0} (\Omega_{0}-1) ^{-\frac{3(1+k)}{2(1+3k)}} \, 
\label{xi}
\end{equation}
and we have assumed $\Omega_0 \approx 1$. Equation~(\ref{messy}) can also be expressed as
\begin{equation}
\Delta_{m} = \Delta_{m}(0)
 \left( 1 - \frac{t_B}{t_m} \right)^2  \left( 1 - \frac{t_B}{t_o} \right)^{-2}\, ,
\label{newdelta}
\end{equation}
where $\Delta_{m}(0)$ is the value of $\Delta_{m}$ when $t_B=0$ (the original expression) and  the last term corresponds to $(H_{ob}/H_{o})^2$ and so effectively cancels the term $(H_o/H_{ob})^2$ in Eq.~(\ref{messy}). The second factor is the important one and 
shows that the overdensity $\Delta_m$ decreases (i.e. $H_{b}L_{\rm max} = \pi \Delta_m^{-1/2}$ increases) as $t_B$ increases. Furthermore, $\Delta_m$ goes to zero  (i.e. $H_{b}L_{\rm max}$ diverges) in the limit $t_B =t_m$ because the background density is then infinite when the overdense region reaches maximum expansion. From Eq.~(\ref{colltime}), 
\begin{equation}
t_m 
= 2^{\frac{1}{1+3k}} \,  \xi  \sqrt{\pi} \; 
\frac{\Gamma \left( \frac{3(1+k)}{2(1+3k)}\right)}{ 
\Gamma \left( \frac{2+3k}{1+3k}\right )} 
\sim  (\Omega_0 -1)^{-\frac{3(1+k)}{2(1+3k)}}(H_b L_{max})^{-1}(1+k)^{-1} \, 
\end{equation}
and this decreases as $k$ increases since all three factors do. 
For any value of $t_B$, this means that there will be a value of $k$
 at which $H_{b}L_{\rm max}$ diverges; 
 $t _m$ falls below $t_B$ for larger $k$, so that $H_{b}L_{\rm max}$ is no longer defined (case c). 
For $t_B<0$, the last factor in Eq.~(\ref{newdelta}) exceeds $1$, so the separate-universe scale is reduced for all $k$. 
This explains the form of the curves in Fig.~\ref{fg:horizon}. The deviations from the earlier form 
essentially arise because the factor $H_b$ is changed.

Although  one cannot apply the above analysis for $t_B > t_m$, this does not mean that one cannot specify a separate-universe scale in this situation -- just that one cannot specify it in terms of the background Hubble scale at maximum expansion.
 After such a positive-curvature regions has attained its maximum expansion, it will begin to recollapse. If it is not too extended, it will virialize at some radius (half its maximum radius in the dust case) due to the developments of pressure gradients and shocks. In this case, 
one can regard the 
region as forming a static spherically symmetric `retarded core' with average density $\rho_c \approx 8 \rho_m$. 
However, this cannot represent a consistent solution of the Einstein equations if the region falls within its Schwarzschild radius and this requires that it be smaller than 
\begin{equation} 
R_{BH} = \left( \frac{3c^2}{8 \pi G \rho_c} \right) ^{1/2} . 
\label{static} 
\end{equation}
The separate-universe scale is somewhat larger than this,
\begin{equation}
L_{SU} = \pi R_{BH} \, ,
\end{equation}
so a black hole again represents an intermediate state before the separate-universe limit.

Finally, the overdense region may collapse {\it before} the big bang (case d), which might be regarded as a separate universe in time rather than space. However, if the collapsing region is just part of a closed universe, it may again virialise at some radius. In this case, it might conceivably persist through  the  background big bang. This possibility is indicated by the broken horizontal line in Fig.~\ref{delayed}(d). However, it is unclear whether this situation is physically realistic, which is why there is a question mark next to this line. 

\section{Separate universe condition for  $k \le -1/3$}

\subsection{$k\to -1/3^+$ limit}
In the limit $k \rightarrow -1/3^+$, Eq.~(\ref{initialtime}) becomes
\begin{equation}
\int^{\eta_{\rm o}}_{0} (1 - \cos \eta)^{\frac{1}{1+3k}} d\eta 
= \left(\frac{1+3k}{2} \right) \left[\frac{2 (\Omega_{\rm o} -1)}{\Omega_{\rm o}}\right] ^{\frac{1}{1+3k}}
\end{equation}
and this goes to $0$ since the above analysis requires that $\Omega_{\rm o}$ be close to $1$,
so that the term in square brackets is less than $1$. 
In the same limit, Eq.~(\ref{colltime}) gives 
\begin{equation}
\int^{\pi}_{0} (1 - \cos \eta)^{\frac{1}{1+3k}} d\eta = 2^{\frac{1}{1+3k}}\sqrt{\pi} \, ,
\end{equation}
which necessarily diverges. Combining these asymptotic behaviours, Eq.~(\ref{eq:Delta_Omega_eta}) gives
 \begin{equation}
\Delta_{m} \approx \frac{ \pi}{(1+3k)^2} \, ,
\end{equation}
and Eq.~(\ref{maximum}) then implies
\begin{equation}
c^{-1}H_{b}L_{\rm max}= \sqrt{\pi} \,  (1+3k) \, ,
\end{equation}
corresponding to a separate-universe scale 
 \begin{equation}
L_{\rm max}  = \frac{1}{2} \sqrt{\pi} (1+3k)^2 ct \, .
\end{equation}
Both of these tend to $0$ as $k \rightarrow -1/3^+$.
This means that for the overdense region to stop expanding,
the density contrast must get very high and hence the curvature 
radius of the overdense region is very small.  
The $k=-1/3$ case itself is clearly very special, because the first term in Eq.~(\ref{growth}) is constant, so we consider that separately.

\subsection{$k=-1/3$}
For $k=-1/3$, the evolution of the positive-curvature Friedmann 
universe is particularly simple.
From Eq.~(\ref{eq:closed_Friedmann}), for $8\pi G\rho_{0}a_{0}^{2}/(3c^{2})=1$, we find a
{\it static} universe with
\begin{equation}
a=a_{0}, \quad  \rho=\rho_{0}.
\end{equation}
Since $L_{\rm max}(t)=\pi a(t)$ is constant, we have
\begin{equation}
c^{-1}H_{b}L_{\rm max}=\frac{\pi a_{0}}{t_{b}},\quad 
\quad \Delta=\frac{\rho}{\rho_{b}}=\frac{\rho_{0}}{\rho_{b0}}\left(\frac{t_{b}}{t_{b0}}\right)^{2} \, ,
\end{equation}
so $c^{-1}H_{b}L_{\rm max}$ decreases and 
the density contrast increases, despite there being
no growing mode in the linear perturbation analysis.
This indicates that such a perturbation must start  nonlinear.

For $8\pi G\rho_{0}a_{0}^{2}/(3c^{2})>1$, Eq.~(\ref{eq:closed_Friedmann}) implies
\begin{equation}
a=a_{0}+\dot{a}_{0}(t-t_{0}),\quad 
\rho=\rho_{0}\left(\frac{a}{a_{0}}\right)^{-2},
\end{equation}
with
\begin{equation}
\dot{a}_{0}=\pm \sqrt{\frac{8\pi G \rho_{0}}{3}a_{0}^{2}-c^{2}} \, .
\label{eq:-1/3dynamical}
\end{equation}
In this case, the universe expands or collapses with a constant speed, depending 
on the initial conditions.
The background Friedmann scale factor and density evolve as 
\begin{equation}
a_{b}=a_{b0}\left(\frac{t_{b}}{t_{b0}}\right),\quad 
\quad \rho_{b}=\rho_{b0}\left(\frac{a_{b}}{a_{0b}}\right)^{-2},
\end{equation}
where $t_{b}$ is the cosmological time since the big bang.
This implies
\begin{equation}
c^{-1}H_{b}L_{\rm max}=\frac{\pi (a_{0}+\dot{a}_{0}(t-t_{0}))}{t_{b}}, \quad 
\quad \Delta=\frac{\rho(t)}{\rho_{b}(t_{b})}=\frac{\rho_{0}}{\rho_{b0}}\left(\frac{a(t)}{a_{b}(t_{b})}\right)^{-2}.
\label{eq:-1/3_dynamical}
\end{equation}
We 
choose $t=t_{b}$, 
corresponding to the synchronous gauge, but the choice of 
$t_{0}$ is still arbitrary. (Note that the $8\pi G\rho_{0}a_{0}^{2}/(3c^{2})<1$ case is prohibited.)

For the expanding case ($\dot{a}_{0}>0$), the density 
contrast $\Delta$ tends to a constant as $t=t_{b}\to \infty$. 
If we choose $t_{0}$ so that the big bang occurs 
simultaneously in the flat background and the
positive-curvature region, which requires $t_{0}=a_{0}/\dot{a}_{0}$, then
$\Delta$ is always constant. This 
is consistent with the linear perturbation theory
and implies
\begin{equation}
c^{-1}H_{b}L_{\rm max}=\pi \dot{a}_{0},
\label{eq:-1/3_expanding_simultaneous_bb}
\end{equation}
where $\dot{a}_{0}$ is the (constant) positive root of Eq.~(\ref{eq:-1/3dynamical}).
Since $\dot{a}_{0}$ is arbitrary, the separate universe length is also arbitrary. 
The assumption of a simultaneous big bang may be unnecessary 
if we consider initial quantum fluctuations which are nonlinear on very 
small scales. 
In this case, although $c^{-1}H_{b}L_{\rm max}$ and $\Delta$ are  initially time-dependent, they tend to constants given by 
Eq.~(\ref{eq:-1/3_expanding_simultaneous_bb})
as time proceeds.
If the perturbed region is sufficiently large but smaller than the 
separate universe scale, this corresponds to a baby universe.
Since it expands to large scales, it can eventually be described classically even though it is 
generated from quantum fluctuations.

For the collapsing case, it is not clear how to choose the origin of 
$t$ with respect to the big bang. In general, $\rho$ begins with 
some finite value and then monotonically increases to infinity. 
One might just exclude this possibility on the grounds that
such a perturbation cannot evolve through the linear  
regime. On the other hand, such a perturbation 
might be generated on very small scales from quantum fluctuations, so we now 
pursue this possibility. 
If we assume 
$a_{0}>0$ and $\dot{a}_{0}<0$ at $t=t_{0}$
with $t_{0}=0$, then we find
\begin{equation}
c^{-1}H_{b}L_{\rm max}=\frac{\pi (a_{0}+\dot{a}_{0}t)}{t}, \quad 
\Delta=\frac{\rho(t)}{\rho_{b}(t)}=\frac{\rho_{0}}{\rho_{b0}}\left(\frac{a_{0}+\dot{a}_{0}t}{a_{b0}t}\right)^{-2} 
\end{equation}
and so $c^{-1}H_{b}L_{\rm max}$ starts infinite and then decreases monotonically to zero at $t=a_{0}/|\dot{a}_{0}|$,
while the density contrast $\Delta$ starts zero and increases monotonically to infinity.
This perturbation is nonlinear from the start and so
can only be generated by quantum fluctuation on very small scales.
If the perturbed region is sufficiently large but smaller than the 
separate universe scale, one has a black hole from the beginning and it is never
classical.

\subsection{$-1<k<-1/3$}
If we try to extend this analysis to the $k < -1/3$ regime, we immediately face the problem that small inhomogeneities decay rather than grow. Therefore the region must have been {\it initially}  collapsing but then bounced into an expansion phase. The density perturbation can only be small well after that, so the `initial' time $t_0$ must also be after the bounce. 
Nevertheless, we can still formally apply the previous analysis by extrapolating back to the time before $t_0$ at which the region was collapsing.

The analysis is different from that in the $k >-1/3$ situation for two
 reasons, both of which follow from Eq.~(\ref{eq:b_tau}). First, $\tau$
 and $t$ have different signs, which means that $\tau$ and hence $\eta$ decrease as $t$
 increases. Second,  $b \propto a^{1+3k}$, with the exponent of $a$
 being negative, so the {\it physical} scale factor $a$ has a qualitatively different
 behaviour from the {\it mathematical} scale factor $b$. 
 The positive-curvature region
 starts off collapsing from a dispersed state, bounces at some finite
 density and then ends up expanding into another dispersed state, so this gives the behaviour of $a(t)$. However, $b$
 starts at $0$ when $\eta = 2 \pi$, increases to a maximum at $\eta =
 \pi$ and then recollapses to $0$ at $\eta = 0$. 
 The crucial point is that the positive-curvature region does not
grow from the background universe through a linear phase. Rather it is like having an extended (non-linear) object in a background universe from the start, analogous to the `retarded core' considered in the $k > -1/3$ case. 

Despite these differences, the separate universe analysis can still be carried out at the bounce epoch (i.e. at the moment of maximum density). Since the scale factor of the perturbed region has a minimum at the bounce, the separate universe scale also has a minimum then.
In this situation, we can no longer take the lower integral limits in
Eq.~(\ref{eq:Delta_Omega_eta}) to be $0$ because the integrals diverge
at this point (i.e. the scale factor $a$ does not become infinite at a
finite time). Instead, we must take the lower limit to be $\eta _B$ with
$0<\eta_{B}<2\pi$, where $\eta_B$ is the value of $\eta$ in the perturbed region corresponding to the big bang time 
of the background universe.
We then replace Eq.~(\ref{eq:Delta_Omega_eta}) with
\begin{equation}
\Delta_m = \left(\frac{\Omega_{\rm o} -1}{\Omega_{\rm o}}\right)
 ^{\frac{3(1+k)}{1+3k}} 
\left[ \frac{\int^{\pi}_{\eta_B} (1-\cos \eta)^{\frac{1}{1+3k}}
 d\eta}{\int^{\eta_{\rm o}}_{\eta_B} (1-\cos  \eta)^{\frac{1}{1+3k}}
 d\eta}\right]^2  \Omega_{\rm o}  \left( \frac{H_o}{H_{ob}} \right)^2\, '
\end{equation}
the first factor being larger than 1 and the second one smaller than 1. 
This expression only applies for $t_B < t_m$ (or $\eta_B > \pi$) since the overdensity is defined with respect to the background universe. Therefore we must assume  $\eta_{0}<\eta_{B}$ and $\pi<\eta_{B}<2\pi$. 
As shown in Appendix B, the integrals have the form
\begin{equation}
\int ^{\pi}_{\eta_B} (1-\cos\eta)^{1/(1+3k)}d\eta \propto [A_k (\eta_B) - A_k(\pi)] 
\end{equation}
and
\begin{equation}
\int^{\eta_0}_{\eta_B} (1-\cos\eta)^{1/(1+3k)}d\eta \propto [A_{k}(\eta_{0})+A_{k}(\eta_{B})-2A_{k}(\pi)] \, ,
\end{equation}
where the function $A_k$ is given by Eq.~(B3).
The separate-universe scale at the bounce therefore becomes
\begin{equation}
c^{-1}H_{b}L_{\rm max}= \pi \Delta_m^{-1/2} = \pi \; \left(\frac{\Omega_{\rm o}}{\Omega_{\rm o} -1}\right)
 ^{\frac{3(1+k)}{2(1+3k)}} 
\left[ 
\frac { 
A_{k}(\eta_{0})+A_{k}(\eta_{B})-2A_{k}(\pi)
} 
{
A_{k}(\eta_{B})-A_{k}(\pi) 
} \right] \, .
\label{eq:Hrmax_negative}
\end{equation}
If we assume that $t_{0}$ is after $t_{\rm m}$ and that  $\eta_{0}$ is sufficiently small for Eq.~(B9) to apply, then
\begin{equation}
A_{k}(\eta_{0})
\approx 2^{3(1+k)/2(1+3k)} \left(\frac{\Omega_{0}-1} {\Omega_{0}}\right)^{3(1+k)/2(1+3k)} \, .
\end{equation}
This term dominates $A_{k}(\eta_{B})$ and $A_{k}(\pi)$ on the right-hand-side of Eq.~(\ref{eq:Hrmax_negative}) and so we obtain
\begin{equation}
c^{-1}H_{b}L_{\rm max}= \frac{2^{3(1+k)/2(1+3k)} \pi }{A_{k}(\eta_{B})-A_{k}(\pi)} \, .
\label{max}
\end{equation}
The dependence on $\Omega_{0}$ disappears but the dependence on $\eta_{B}$ remains. Using Eq.~(B6) for $A_k(\pi)$, 
we finally obtain
\begin{equation}
c^{-1}H_{b}L_{\rm max} 
 =  2 \sqrt{\pi}\;  \frac{(1+3k)}{3(1+k)} \frac{
\Gamma \left( \frac{2+3k}{1+3k}\right )}{ \Gamma \left( \frac{3(1+k)}{2(1+3k)}\right)} \left[ \frac{A_{k}(\eta_{B})}{A_{k}(\pi)} -1 \right]^{-1}\, .
\end{equation}
Apart from the factor in square brackets, this is the same as in the $k>-1/3$ case. The first factor implies $L_{\rm max} \rightarrow 0$ as $k \rightarrow -1/3^-$ and $L_{\rm max} \rightarrow \infty$ as $k \rightarrow -1$.  Although the numerator diverges if $(2+3k)/(1+3k)$ is a negative integer,  corresponding to $k = -2/3, -1/2, -4/9..... -1/3$, Eq.~(\ref{max}) implies that
the total expression is finite because $A_{k}(\eta_{B})-A_{k}(\pi)$ is finite for $-1<k<-1/3$.   
The form of of $c^{-1}H_{b}L_{\rm max}$ for various values of $\eta_B/\pi$ in the range $1$ to $2$ (i.e. for $t_B < t_m$) is shown in Fig.~\ref{fg:horizon}. 
Divergence occurs at $\eta_B = \pi$ (i.e. $t_m = t_B$) and this corresponds to the limit $k=-1/3$. 

\section{Discussion}

Prompted in part by the work of Kopp {\it et al.}, we have addressed
various inadequacies in our previous treatment of the separate-universe
problem. (1) We have compared our analysis with a more rigorous
curvature perturbation approach. (2) We have allowed for the necessity of a compensated inhomogeneity in order to preserve the flat FRW background. (3) We have extended the analysis to include all equations of state with $-1 < k < \infty$ and find that the interpretation is very different for $k<-1/3$ and $k>-1/3$.  
(4) We have analysed the problem in terms of the Hubble horizon rather
than the  particle horizon and distinguished between the background and
local horizons. 

There are still several issues to be resolved. (1) We need to
understand the effects of the density gradients in a more realistic
situation. (2) We should relate our analysis to the self-similar black
hole and wormhole solutions. (3) We should understand the
significance of the self-similar universal black hole solutions (which
have an apparent horizon but no event horizon). (4) We need to revisit
the problem of black hole formation and consider the nature of the Jeans
condition for $k < -1/3$. (5) We should determine whether a region
bigger than the Jeans length is necessarily a type II fluctuation. (6) We should study the Penrose and embedding diagrams for these solutions carefully and relate this to the work of KHW.
(7) In order to distinguish between separate univeres, black holes, wormholes and baby universes, we must carefully analyse the behaviour of the trapped surfaces in these solutions and this problem is addressed in a separate paper \cite{carrharada2}. 

The separate-universe condition can never be attained in any natural context (such as inflation) since the topology can never change in classical relativity but one might worry about it in the quantum gravitational context. Although the separate-universe condition does not explicitly affect the PBH mass distribution, it does implicitly since the maximum mass of a PBH forming at any epoch is simply related to the separate-universe scale.

\acknowledgments
BC and TH  are grateful to the Royal Society and JSPS for an exchange visitors grant. BC also thanks RESCEU, University for Tokyo, and TH thanks the Astronomy
Unit, Queen Mary, University of London, for hospitality received during this work.
TH was supported by the Grant-in-Aid for Young Scientists (B) (No. 21740190)
and Challenging Exploratory Research (No. 23654082) for Scientific
Research Fund of the Ministry of Education, Culture, Sports, Science and 
Technology, Japan and Rikkyo University Special Fund for Research.

\appendix

\section{Analysis of Kopp {\it et al.}}

Here we compare our approach with the analysis of the separate-universe problem by KHW \cite{Kopp:2010sh}.  We first emphasize the points on which we agree.
 It is clearly sensible to describe the situation in terms of the curvature fluctuation ($\zeta$) rather than the density fluctuation ($\delta$), although we would claim that this is not a fundamental distinction. We also agree that it is important to distinguish between measuring the size of an overdense sphere in terms of the proper radial distance ($L$) and the areal radial distance ($R$). Indeed, we use their notation in the present paper  to allow a direct comparison of our results and end up with a separate-universe condition corresponding almost  exactly with theirs. One strength of the present work is that we allow the equation-of-state parameter $k$ to vary over all possible values, including 
$k<-1/3$, whereas the KHW analysis is confined to the dust ($k=0$) and radiation  ($k=1/3$) cases. 

KHW describe the overdensity in terms of the curvature perturbation $\zeta (r,t)$, which is defined in terms of the perturbation to the spatial part of the metric (\ref{closedmetric}) when expressed in the conformally flat form:
\begin{equation}
^{(3)}ds^2 = b(t)^2 e^{2\zeta (r,t)}[dr^2+r^2 (d\theta^{2}+\sin^{2}\theta)]  \, .
\end{equation}
They introduce three measures of this: a {\it central} volume fluctuation $\zeta(0,t)$, an {\it averaged} volume fluctuation $\overline{\zeta}(t)$, and the density fluctuation when the region falls within the Hubble horizon $\delta_{KHW}$. All three  quantities can be expressed in terms of the extent of the overdense region $\chi_a$. In the radiation case, they find   
\begin{equation}
\zeta \approx -2 \ln \cos \frac{\chi_a}{2}, \quad \overline{\zeta} \approx \frac{1}{3} \ln \frac{2(\chi_a - \sin \chi_a \cos \chi_a)}{2 \sin^3 \chi_a}, \quad
\delta_{KHW} = \frac{1}{16} \sin^2\chi_a(8 + \sin^2 \chi_a) \, ,
\label{zetas}
\end{equation}
and these three functions are plotted in their Figure (4). 
The first two expressions  apply only so long as the region is much larger than the Hubble horizon; otherwise there are gradient effects and $\zeta$ and  $\overline{\zeta}$ become time-dependent. Equation~(\ref{zetas}) shows that $\zeta$ and $\overline{\zeta}$ diverge, while $\delta_{KHW}$ goes to zero, as the size of the region tends to the separate-universe scale ($\chi_a \rightarrow \pi$). KHW infer that the separate-universe condition can never be satisfied but there is still a separate-universe scale and the fact that $\zeta$  and $\overline{\zeta}$ diverge there merely indicates that something strange happens and not that this condition cannot be attained. Note that $\delta_{KHW}$ is different from what we term $\delta_H$, which can be shown to be equivalent to  $4\zeta$ if $\chi_a \ll 1$.

KHW differentiate between Type I and Type II fluctuations 
but it is important to appreciate that one gets PBHs even in the type-II situation. It is interesting that the final PBH mass corresponding to $\chi$ and  $\pi  - \chi$ are the same and KHW usefully  clarify what happens as $\chi \rightarrow \pi$, showing that $\delta \rightarrow 0$ in the limiting case. 
As $\chi$ increases, the Misner-Sharp mass reaches a maximum at $\chi = \pi/2$ and then shrinks to $0$ as $\chi \rightarrow \pi$. This is why they claim that the separate-universe condition is necessarily avoided. They infer that one cannot place any constraint on the form of $\delta_0$ and that the strongest constraint on the fluctuations comes from $\chi = \pi/2$ rather than $\chi = \pi$.  We agree with this but  in our view the fact that the mass tends to $0$ as one approaches the separate-universe condition is just an equivalent way of stating the problem, albeit  an illuminating one. 

\section{Hypergeometric function}

The indefinite integrals of interest all have the form
\begin{equation}
\int ^{\eta} (1-\cos\eta)^{1/(1+3k)}d\eta = \pm \frac{\sqrt{2}(1+3k)}{3(1+k)}A_{k}(\eta)
\end{equation}
where 
\begin{equation}
A_{k}(\eta)\equiv (1-\cos\eta)^{3(1+k)/2(1+3k)}
F\left(\frac{1}{2},\frac{3(1+k)}{2(1+3k)};\frac{5+9k}{2(1+3k)};\frac{1-\cos\eta}{2}\right)
\label{solution}
\end{equation}
and $F(\alpha,\beta;\gamma;x)$ is the hypergeometric function.
The positive sign applies for $0 < \eta < \pi$ and the negative sign for $\pi < \eta < 2\pi$. We can rewrite the last expression in the form
\begin{equation}
A_k(\eta)  = (1-\cos\eta)^{\beta}
F \left(\frac{1}{2}, \beta; \beta + 1; \frac{1-\cos\eta}{2}\right) \, ,
\end{equation}
where $ \beta  \equiv  3(1+k)/2(1+3k)$. To evaluate the integral in the limits $\eta \ll 1$ (as required in the $k > -1/3$ case) and $\eta = \pi$ (as required in all cases), we recall the general result:
\begin{equation}
F(\alpha,\beta;\gamma;0)=1, \quad F(\alpha,\beta;\gamma;1)=\frac{\Gamma(\gamma)\Gamma(\gamma-\alpha-\beta)}{\Gamma(\gamma-\alpha)\Gamma(\gamma-\beta)}.
\label{hyper}
\end{equation}
In this case, $\alpha = 1/2$ and $\gamma = \beta + 1$, so we can rewrite the second relation as 
\begin{equation}
F(1/2,\beta;\beta+ 1;1)=\frac{\Gamma(\beta + 1) \Gamma(1/2)}{\Gamma(\beta +1/2)\Gamma(1)} = \frac{\sqrt{\pi} \, \Gamma(\beta+1)}{\Gamma (\beta +1/2)} 
\end{equation}
\if
The separate universe scale can then be expressed as
\begin{equation}
c^{-1}H_{b}L_{\rm max}= \pi \Delta^{-1/2} = \pi \; \left(\frac{\Omega_{\rm o}}{\Omega_{\rm o} -1}\right)
 ^{\frac{3(1+k)}{2(1+3k)}} 
\left[ 
\frac { 
A_{k}(\eta_{0})+A_{k}(\eta_{B})-2A_{k}(\pi)
} 
{
A_{k}(\eta_{B})-A_{k}(\pi)
} \right] \, 
\label{eq:Hrmax_negative}
\end{equation}
at the bounce.
This term dominates $A_{k}(\eta_{B})$ and $A_{k}(\pi)$ on the right-hand-side of Eq.~(\ref{eq:Hrmax_negative}) and so we obtain
\begin{equation}
c^{-1}H_{b}L_{\rm max}= \frac{2^{3(1+k)/2(1+3k)} \pi }{A_{k}(\eta_{B})-A_{k}(\pi)}.
\label{max}
\end{equation}
The dependence on $\Omega_{0}$ disappears but the dependence on $\eta_{B}$ remains. 
\fi
and express $A_{k}(\pi)$ in terms of gamma functions:
\begin{equation}
A_{k}(\pi)=\sqrt{\pi} \, 2^{3(1+k)/2(1+3k)} \frac{\Gamma\left(\frac{5+9k}{2(1+3k)}\right)}
{\Gamma\left(\frac{2+3k}{1+3k}\right)}   \, .
\end{equation}
Although the two $\Gamma$ terms diverge at particular values of $k$, their divergences cancel and the total expression is always finite. Equation (B1) then implies
\begin{equation}
\int ^{\pi} (1-\cos\eta)^{1/(1+3k)}d\eta 
 = \pm \, 2^{\frac{1}{1+3k}} {\sqrt \pi} \,  \frac{\Gamma\left(\frac{3(1+k)}{2(1+3k)}\right)}{\Gamma\left(\frac{2+3k}{1+3k}\right)} \, ,
\end{equation}
where we have simplified the expression using the relation $\Gamma(\beta + 1) = \beta \, \Gamma(\beta)$, which applies unless $\beta =0$.

We also need an expression for $A_k(\eta_o)$ for $\eta_o \ll 1$, corresponding to some time when the density perturbation is small. In this case, the second order approximation
\begin{equation}
\cos^{-1} x \approx (6 - 2 \sqrt{3 + 6x}  \, )^{1/2} \quad \mathrm{with} \quad x = \frac{2-\Omega_o}{\Omega_o}
\end{equation}
leads to Eq.~(\ref{eq:eta_Omega}). We also have
\begin{equation}
\int ^{\eta_o} (1-\cos\eta)^{1/(1+3k)}d\eta \approx  \int ^{\eta_o} \left(\frac{\eta^2}{2}\right)^{1/(1+3k)} \left[ 1- \frac{\eta^2}{12(1+3k)} \right] d\eta \, ,
\end{equation}
so integrating and using Eq.~(\ref{eq:eta_Omega}) gives Eq.~(\ref{initialtime}). The first term also comes from Eq.~(\ref{solution}):
\begin{equation}
A_{k}(\eta_{0})\approx \left(\frac{\eta_{0}^2}{2}\right)^{\beta} F \left(\frac{1}{2}, \beta; \beta + 1; 0 \right) =  \left(\frac{\eta_{0}^2}{2}\right)^{3(1+k)/2(1+3k)}  \, ,
\end{equation}
which is small for $k > -1/3$ but large for $k < -1/3$. Special consideration needs to be given to the $k = -1/3$ and $k = -1$ cases.


\begin{thebibliography}{99}

\bibitem{ch1974}
   B.J.~Carr and S.W.~Hawking,
   Mon. Not. R. Astron. Soc. {\bf168}, 399 (1974).

\bibitem{carr1975}
B.J.Carr, Astrophys. J. {\bf 201}, 1-19 (1975).
\bibitem{hc2005c}
T.~Harada and B.J.~Carr,
Phys. Rev. D{\bf 71}, 104010 (2005).

\bibitem{nnp1978}
D.~K.~Nadezhin, I.~D.~Novikov and A.~G.~Polnarev,
Sov. Astron.
{\bf 22}, 129 (1978).

\bibitem{mp2009}
I.~Musco and A.G.~Polnarev,
Class. Quant. Grav.
{\bf 24}, 1405 (2009).

\bibitem{bardeen}
J.~Bardeen, Phys. Rev. D{\bf 22}, 1882 (1980).

\bibitem{Kopp:2010sh}
  M.~Kopp, S.~Hofmann and J.~Weller,
  Phys.\ Rev.\ D{\bf 83} 124025 (2011).
 
\bibitem{supernova} 
S.~Perlmutter {\it et al.},
Astrophys.\ J.\  {\bf 517}, 565 (1999);
A.G.~Riess {\it et al.},
Astron.\ J.\  {\bf 116}, 1009 (1998);
Astron.\ J.\  {\bf 117}, 707 (1999).

\bibitem{ratra} 
B.~Ratra and J.~Peebles, Phys. Rev. D{\bf 52}, 1837 (1995).


\bibitem{chm2010}
B.J.~Carr, T.~Harada and H.~Maeda, 
Class. Quant. Grav. {\bf 27}, 183101 (2010).

\bibitem{hmc2007}
T.~Harada, H.~Maeda and B.J.~Carr, Phys. Rev. D{\bf 77}, 024022 (2008).

\bibitem{mhc2007}
H.~Maeda, T.~Harada and B.J.~Carr, Phys. Rev. D{\bf 77}, 024023 (2008).

\bibitem{mhc09}
H.~Maeda, T.~Harada and B.J.~Carr, 
Phys. Rev. D{\bf 79}, 044034 (2009).

\bibitem{carrharada2}
B.J.~Carr and T.~Harada, preprint (2014).


\bibitem{jedamsik}
J.~Niemeyer and K.~Jedamzik, 
Phys. Rev. D{\bf 
59}, 124013 (1999). 

\bibitem{green}
A.M.~Green, A.R.~Liddle, K.A.~Malik and M.~Sasaki, 
Phys. Rev. D{\bf 
70}, 041502 (R) (2004). 

\bibitem{musco}
I.~Musco, J,~Miller and L.~Rezzolla, Class. Quant. Grav. {\bf 
22}, 1405 (2005).

\bibitem{Harada:2013epa} 
  T.~Harada, C.~M.~Yoo and K.~Kohri,
  Phys.\  Rev.\ D{\bf 88}, 084051 (2013).

\bibitem{hayward1996}
S.A.~Hayward, 
Phys. Rev. D{\bf 53}, 1938 (1996). 

\bibitem{faraoni}
V.~Faraoni, arXiv:1106.4427 (2011).

\bibitem{hidalgo}
J.S.~Bullock and J.R.~Primack, Phys. Rev. D{\bf  55} 7423 (1997); 
P.~Ivanov, Phys. Rev. D{\bf  57} 7145 (1998);
 J.C.~Hidalgo, arXiv:0708.3875 (2007); 
C.T.~Byrnes, E.J.~Copeland and  A.M.~Green, Phys. Rev. D{\bf  86} 043512 (2012).

\end{thebibliography}
\end{document}